\documentclass{cpbtex}
\usepackage{CJKutf8}
\usepackage{graphicx,xcolor}
\usepackage{subfig}
\usepackage{hyphenat}
\usepackage{aasmacros}
\usepackage{xspace}
\usepackage{url}
\usepackage{amsmath}
\usepackage{hyperref}
\usepackage{url}
\usepackage{floatrow}
\usepackage[framemethod=tikz]{mdframed}
\usepackage{caption}
\usepackage{subfig}
\usepackage{academicons}
\usepackage{environ}
\usepackage{soul}

\newcommand{\orcid}[1]{\href{https://orcid.org/#1}{\includegraphics[width=8pt]{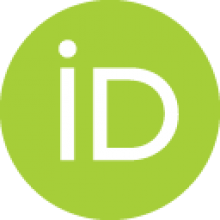}}}

\newcommand\snrb{${\rm SNR_{B}}$}
\newcommand\snrr{${\rm SNR_{R}}$}
\newcommand\bprp{$G_{\rm BP} - G_{\rm RP}$}
\newcommand\MG{M_\textrm{G}}

\newcommand{\kms}{${\rm kms^{-1}}$}
\newcommand\sigmarv{$\sigma_{\rm rv}$}
\newcommand\sigmarvb{$\sigma_{\rm rv}^{\rm B}$}
\newcommand\sigmarvr{$\sigma_{\rm rv}^{\rm R}$}
\newcommand\nvisitt{$N_{\rm vist} \geq 2$}

\title{LAMOST Medium-Resolution Spectroscopic Survey of Binarity and Exotic Star (LAMOST-MRS-B): Observation Strategy and Target Selection}
\begin{document}
\begin{CJK*}{UTF8}{gbsn}

%\date{\today}
\maketitle{}

\author{Jiao Li(李蛟)\orcid{0000-0002-2577-1990}$^{1}$, \and Jiang-Dan Li(李江丹)\orcid{0000-0003-3832-8864}$^{2,3}$, \and
        Yan-Jun Guo(郭彦君)\orcid{0000-0001-9989-9834}$^{2,3}$, \and Zhan-Wen Han(韩占文)\orcid{0000-0001-9204-7778}$^{2}$, \and Xue-Fei Chen(陈雪飞)\orcid{0000-0001-5284-8001}$^2$, \and Chao Liu(刘超)$\orcid{0000-0002-1802-6917}^{1}$, \and
        Hong-Wei Ge(葛宏伟)$^2$, \and
        Deng-Kai Jiang(姜登凯)$^2$, \and Li-Fang Li(李立芳)$^2$, \and
        Bo Zhang (章博)\orcid{0000-0002-6434-7201}$^{1}$, \and Jia-Ming Liu(刘佳明)$^{5}$, \and
        Hao Tian(田浩)\orcid{0000-0003-3347-7596}$^{1}$, \and Hao-Tong Zhang(张昊彤)$^{4}$, \and
        Hai-Long Yuan(袁海龙)$^{4}$, \and Wen-Yuan Cui(崔文元)$^{5}$, \and
        Juan-Juan Ren(任娟娟)$^{1}$, \and
        Jing-Hao Cai(蔡靖豪)$^{2,3}$ and
        Jian-Rong Shi (施建荣)$^{4}$

% The Chinese charactics: 李蛟, 李江丹, 郭彦君, 陈雪飞, 刘超, 葛宏伟, 姜登凯, 李立芳, 章博, 刘佳明, 田浩, 张昊彤, 袁海龙，崔文元, 蔡靖豪, 韩占文

\thanks{Corresponding author. E-mails: lijiao@bao.ac.cn; liuchao@nao.cas.cn}\\
$^{1}${Key Laboratory of Space Astronomy and Technology, National Astronomical Observatories Chinese Academy of Sciences, Beijing 100101, People's Republic of China}\\  % The line break was forced via \\
$^{2}${Yunnan observatories, Chinese Academy of Sciences, P.O. Box 110, Kunming, 650011, China}\\
$^{3}${School of Astronomy and Space Science, University of Chinese Academy of Sciences, Beijing, 100049, People's Republic of China}\\
$^{4}${Key Laboratory of Optical Astronomy, National Astronomical Observatories, Chinese Academy of Sciences, Beijing 100101, People's Republic of China}\\
$^{5}${Department of Physics, Hebei Normal University, Shijiazhuang 050024, People's Republic of China}\\ % The line break was forced via \\
}   % The line break was forced via \\

 %
% 1. For Chinese authors, the name in Chinese characters should also be given. For example, Gang Liu(Áõ¸Õ), Xiao-Ming Li(ÀîÏþÃ÷)
% 2. Please ensure that every author approves the submission of the manuscript
% 3. Abbreviations should not be used in the affiliations

\begin{abstract}
LAMOST-MRS-B is one of the sub-surveys of LAMOST medium-resolution ($R\sim7500$)
spectroscopic survey. It aims at studying the statistical properties (e.g. binary fraction, orbital period distribution, mass ratio distribution) of binary stars and exotic stars. 
We intend to observe about 30,000 stars ($10\leq G < 14.5$ mag) 
with at least 10 visits in five years. 
We first planned to observe 25 plates around the Galactic plane in 2018. Then the plates were reduced to 12 in 2019 because of the limitation of observation. At the same time, two new plates located at the high Galactic Latitude were added to explore binary properties influenced by the different environments. In this survey project, we set the identified exotic and low-metallicity stars with the highest observation priorities. For the rest of the selected stars, we gave the higher priority to the relatively brighter stars in order to obtain high quality spectra as many as possible. Spectra of 49,129 stars have been obtained in LAMOST-MRS-B field and released in DR8, of which 28,828 and 3,375 stars have been visited more than twice and ten times with $\rm SNR \geq 10$, respectively. Most of the sources are B-, A-, and F-type stars with $\rm -0.6< [Fe/H] < 0.4 $ dex. We also obtain 347 identified variable and exotic stars and about 250 stars with $\rm [Fe/H] <-1$ dex. We measure radial velocities (RVs) by using 892,233 spectra of the stars. The uncertainties of RV achieve about $1$ and $10$ \kms for 95\% of late- and early-type stars, respectively. The  datasets presented in this paper is available in \href{https://www.scidb.cn/en/s/222q6j}{Science Data Bank}\footnote{\url{http://www.doi.org/10.57760/sciencedb.j00113.00035}}.
\end{abstract}

%\href{https://www.scidb.cn/en/s/222q6j}
\textbf{Keywords:} Surveys, Spectroscopy, Catalogs, Binary%no more than four sets of keywords should be provided

\textbf{PACS:} 95.80.+p, 95.75.Fg, 95.80.+p, 97.80.-d

%\textbf{PACS:} no more than four \href{http://cpb.iphy.ac.cn/EN/column/item208.shtml}{PACS codes} should be provided

\section{Introduction} \label{sec:intro}

Binary stars are common in the Galaxy. In general, more than 50\% of stars are believed to be in form of binaries or multiple systems\cite{Duchene2013ARA&} (hereafter, we abbreviate binary and multiple system as binary unless otherwise specified). The binary fraction increases with stellar mass\cite{Duchene2013ARA&, Moe2017ApJS, Offner2022arXiv}. For solar-type stars with masses around 0.8 - 2 $M_\odot$, the binary fraction is not less than 50\%\cite{Fuhrmann2012ApJs, Tokovinin2014AJI}, while for more massive stars ($>8~M_{\odot}$), the binary fraction may exceed 70\%\cite{Sana2012Sci}.  Binaries with orbital periods less than a few thousand days, would experience mass transfer during their evolution\cite{Hanzhanwen2020RAA, Moe2017ApJS}. 
For example, symbiotic stars are transferring mass from the giant components to the compact components; their orbital periods range from hundreds of days to a few tens of years\cite{Mohamed2012BaltA, Mikoajewska2012BaltA}.
Meanwhile, over 70\% of all massive stars will exchange mass with a companion\cite{Sana2012Sci}.
We identify a close binary as the binary star would experience mass transfer for clarity.

The presence of a nearby companion will alter the stellar evolution fate. It is widely known that there are exotic objects e.g. blue stragglers, barium stars, symbiotic stars, double degenerates, hot subdwarf stars, cataclysmic variables (CVs), pulsars, X-ray binaries, gamma-ray bursts, Type Ia supernovae (SNe Ia) and double stellar black holes (BHs) which are led by the interaction of binary stars\cite{Sana2012Sci, Hanzhanwen2020RAA, Belczynski2020ApJ}. However, several crucial processes are still puzzled in the binary evolution such as mass and angular momentum loss, dynamical mass transfer and common envelope evolution. Searching for exotic objects will help us to understand how the exotic objects form and what the ultimate fate of the object is.

Stellar structure and evolution are one of the cornerstones to study galaxies. There are some strange phenomena in galaxies that are hardly explained by single stellar population synthesis (SPS). For example, the far-ultraviolet (FUV) excess in the spectra of elliptical galaxies\cite{Burstein1988ApJ}, while it can be well reproduced by including the radiation of hot subdwarf stars (sdOB) resulting from binary interactions\cite{Han2007MNRAS}. The accretion of white dwarfs (WDs) in binary stars could interpret the existence of soft X-ray band extended emissions in galaxies\cite{Chenhailiang2015MNRAS}. It has a big impact on deriving ages, masses and star formation rates of galaxies by considering the binary interaction in SPS --- bianry population synthesis (BPS) \cite{Hanzhanwen2020RAA}. The cosmic re-ionization photons based on SPS models are less than the required amount by a few factors, but BPS models provide a sound physical basis for cosmic re-ionization\cite{Secunda2020ApJ, Gotberg2020A&A}. In BPS study, binary samples are generated based on the parameters: the star formation rate, the initial mass function of the primary, and the statistical properties of binaries such as the mass ratio distribution, the orbital period distribution and the orbital eccentricity distribution. The evolution result of BPS model would be different when the recipes of the initial parameters are changed. The final binary properties would also vary with the evolution. Therefor, the initially parameters can be constrained by comparing the observed statistical properties of binary with BPS models.

The statistical properties of close binary vary with the stellar mass, and metallicity. Offner et al. (2022) summarized that the close binary fraction increases from $\sim20\%$ to $\sim70\%$ as the primary stars move from $\sim0.1$ to $\sim30 M_{\odot}$ by compiling the results of observational and theoretical studies of stellar multiplicity\cite{Offner2022arXiv}. The orbital period (or separation) distributions are different between massive stars and solar-type (and the smaller mass) stars. The orbital period distribution of massive stars is likely to follow Opik's law\cite{Sana2014ApJS, Moe2017ApJS}, while the distribution of smaller mass satisfies lognormal distribution\cite{Raghavan2010ApJS, Winters2019AJ}. And the peaks of lognormal distribution increase with the primary mass, the peak reaches to $\log P/({\rm day}) = 5.03$ for the solar-type stars. 

Assuming the distribution of mass ratio follows a power law, the power index ($\gamma$) decreases with the mass of primary star\cite{Moe2017ApJS}. In addition, there is an excess of twins ($q>0.95$) for the distribution of solar-type and smaller mass stars\cite{Duchene2013ARA&, Kobulnicky2016, Moe2019ApJ}, and the excess twins fraction might arrive to 30\%\cite{Moe2017ApJS}. Nonetheless, the excess twin fraction among massive stars is only 10\% within SMA $a < 0.5$ au, and the excess cannot be detected among O-type binaries beyond $a > 0.5$ au at all\cite{Sana2012Sci, Offner2022arXiv}.

The relation between binary fraction ($f_{b}$) and metallicity has been widely studied. A few works claim that $f_b$ is not correlated\cite{Latham2002AJ, Moe2013ApJ} or positively correlated\cite{Carney1983AJ, Abt1987ApJ, Hettinger2015ApJ} with metallicity. However, with the released data of Apache Point Observatory Galactic Evolution Experiment (APOGEE)\cite{Majewski2017AJ} survey and the Large Sky Area Multi-Object fiber Spectroscopic Telescope (LAMOST) survey\cite{Cuixiangqun2012RAA, Zhaogang2012RAA}, $f_b$ is found to be anti-correlated with metallicity\cite{Gaoshuang2014ApJ, Gaoshuang2017MNRAS, Tianzhijia2018RAA, Badenes2018ApJ, Moe2019ApJ, Mazzola2020MNRAS}. The relation looks like to change with primary mass as well. Raghavan et al. (2010) found no correlation between stellar companions and metallicity for $B - V < 0.625$, but among the redder subset (the stellar mass is smaller), metal-poor stars are more likely to have companions\cite{Raghavan2010ApJS}. A similar result was also found by Liu (2019), that the $f_b$ is quite flat with metallicity as $\gamma < 1.2$, while the relation is anti-correlated as $\gamma > 1.2$ (primary mass is smaller)\cite{Liuchao2019MNRAS}. The eccentricity distribution is a function of the period. The power-law exponent of the eccentricity distribution is $\eta = -0.3 \pm 0.2$ for $p = 3-10$ days, while is $\eta = 0.6\pm0.3$ for $p = 10-500$days\cite{Moe2017ApJS}. These indicate that the distributions of binary orbital parameters are not independent.

The clear correlation between the distributions is ambiguous, due to the absence of larger enough binary samples and the different selection bias from the current observation programs. It is not clear whether these distributions are universal, since most of works were investigated based on neighborhood stars within 100 pc. 

Interaction of close binary stars would generate some peculiar stars and could influence the result of BPS, so that we pay more attention to the orbital parameter distributions of close binary. After the first stage 5-year survey of LAMOST low resolution spectra (LAMOST-LRS), a median resolution spectral survey is initiated (LAMOST-MRS). As a sub-survey of LAMOST-MRS, MRS-B will provide the largest homogeneous sample with multiple observations to study the stellar multiplicity, discover more various exotic binaries under different evolutionary stages, and measure their orbital parameters.

\section{LAMOST-MRS}

LAMOST is a 4-meter quasi-meridian reflective Schmidt telescope with 4000 fibers installed on its 5- degree-FoV focal plane. It has achieved more than 9 million stellar spectra with spectral resolution of $R \sim 1800$ at the first stage survey from 2011 to 2018. LAMOST has appended 16 spectrographs with spectra resolution of $R\sim7500$ during 2017\cite{Liuchao2020arxiv}. A pilot program of LAMOST-MRS was performed from September 2017 to June 2018 accompanied by LAMOST-LRS. LASMOT-MRS was being executed regularly from October 2018 to June 2023, and will take about 50\% nights to observe objects. The spectrographs will be switched to medium resolution mode from the 7th until the 22nd of each lunar month. 

\begin{figure}
    \centering
    \makebox[\textwidth][c]{\includegraphics[width=\textwidth]{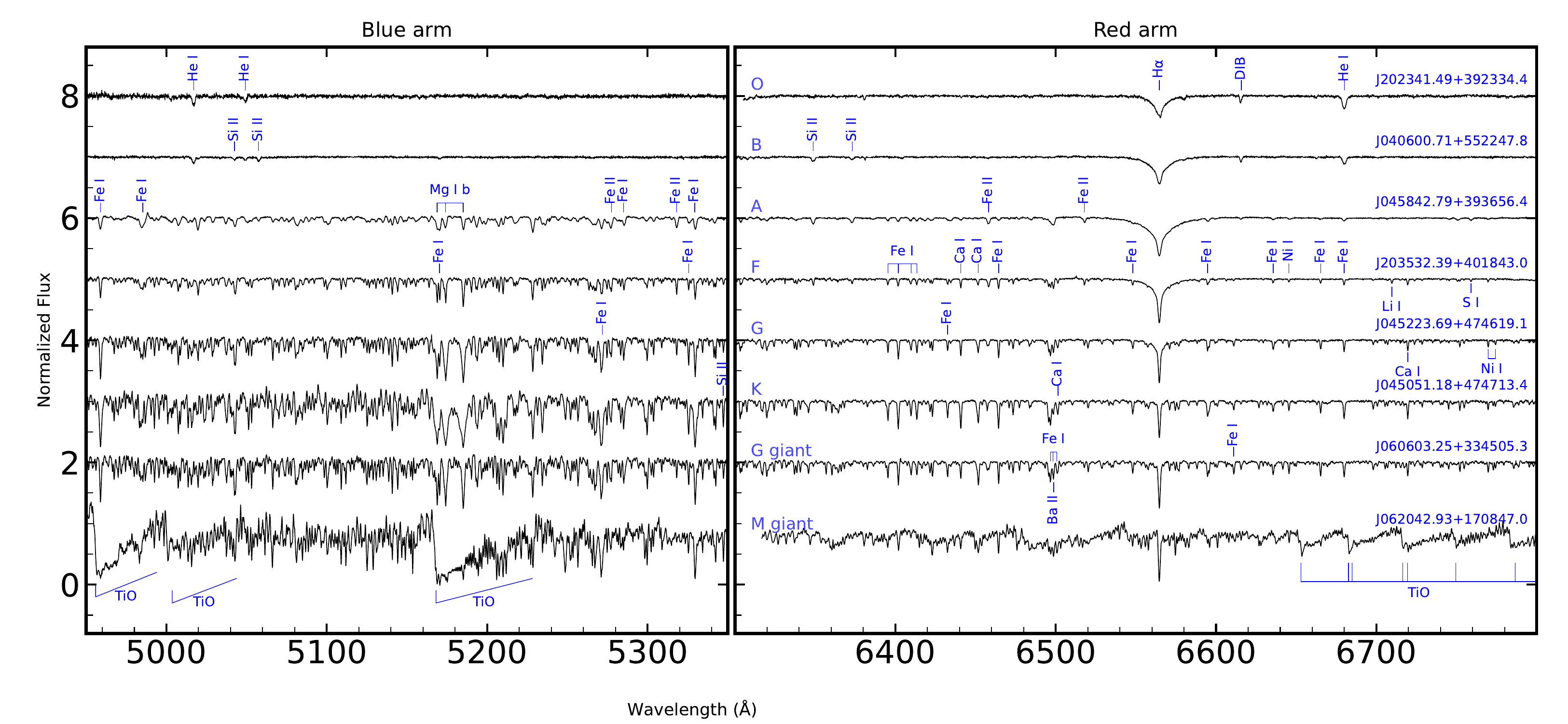}}
    \caption{The spectra example of LAMOST-MRS from late O- to M-type stars selected from LAMOST-MRS-B randomly. The spectra are normalized by following the process of Zhang et al. (2021)\cite{Zhangbo2021ApJS}. Wavelengths of the spectra have been corrected to rest frame with the corresponding radial velocities. DIB is diffuse interstellar bands (6614 \r{A}). The wavelength of TiO bands can be found in Table 6 of Valenti et al. (1998)~\cite{Valenti1998ApJ}. The wavelength of the atomic absorption lines can be found from \href{https://physics.nist.gov/PhysRefData/ASD/lines_form.html}{NIST Atomic Spectra Database Lines Form.
    }
    }
    \label{fig:spec}
\end{figure}

Different with the wavelength coverage of LAMOST-LRS, LAMOST-MRS covers two parts of the spectra, see Figure~\ref{fig:spec}, the blue [4950, 5350 ]~\r{A} and the red [6300, 6800]~\r{A} arms. For late-type stars (F-, G-, and K-type), the blue arm spectra deliver the information about 20 elemental abundances, including Li, C, Na, Mg, Si, Ca, Sc, Ti, V, Cr, Mn, Fe, Co, Ni, Cu, Ba, Y, Sm, and Nd. With those information, the radial velocities (RVs) can be constrained with precision of about 1 \kms. While for the early-type (O-, B-, and A-type) stars, the absorption lines are weak and sparse in the blue arm\cite{Liuchao2020arxiv}. So it is better to measure RV by using the spectra of the red arm with the precision of 10 \kms. In the meantime, the red arm spectra covers ${\rm H\alpha}$, [NII], and [SII] emission lines, which are crucial to the study the Galactic nebulae, such as the HII regions, the supernova remnants, and the planetary nebular, etc. The ${\rm H\alpha}$ profile is very useful in researching on the proto-planetary disk surrounding the very young stars as well.

LAMOST-MRS is separated as time-domain (TD) and non time-domain (NT) surveys. It scientific goals mainly include stellar multiplicity, stellar pulsation, host stars of exoplanets, star formation, emission nebulae, Galactic archaeology, and open clusters. According to the scientific goals, LAMOST-MRS is divided into MRS-B (TD), MRS-K (TD)\cite{Fujianning2020RAA, Zongweikai2020ApJS}, MRS-T (TD), MRS-S (TD or NT), MRS-N (NT)\cite{Wuchaojian2021RAA}, MRS-G (NT) fields and  MRS-O (NT)\cite{Liuchao2020arxiv}. Each field has a PI who is responsible for the target selection. For the TD survey, a plate is observed with 3-8 single 1200 seconds exposures in each visit (visit is that a plate is observed in one night, no matter how many exposures it takes) within the  4-hour observation window of LAMOST. About 60\% visits are assigned to the TD survey. For the NT survey, a plate is continuously observed three single 1200 seconds exposure in each visit.

In this paper, we mainly focus on introducing the target selection and observation strategies of MRS-B, which aims at stellar multiplicity and searching for exotic stars. We have $\sim 30$\% observation time in the TD survey. A ``planid'' is assigned to each LAMOST-MRS plate (pointing), and the planids of MRS-B are labeled as ``TDhhmmssNddmmssB01'', where ``hhmmssNddmmss'' denotes the equatorial coordinates of the direction center star of each plate, ``N'' means north.

\section{Observation Strategy}

LAMOST-MRS are designed to observe about 4,000 targets simultaneously. About 3,000 stars can be observed  excluding a few problematic fibers and a few hundreds of sky fibers used to obtain the skylight. There are not enough bright sources ($G < 10$ mag) to occupy all the 3,000 fibers. When observing the bright sources ($G < 10$ mag), the CCD will be saturated if the same exposure time is taken as the faint source ($G \sim 14$ mag). On the other hand, if the sources are too faint, we need to take a longer exposure time to obtain spectra with a good signal-to-noise ratio (SNR), which would smear the variations of short period stars and induce more contamination from cosmic rays. Limited the designation of LMAOST, the observation window for each target is only four hours, it means that the longer exposure time would reduce the number of exposure. Consequently, the variation of some variable stars with periods of a few hours such as RR Lyrae\cite{Marconi2011ApJ,Jurcsik2017MNRAS} will be lost. Based on the experience of LAMOST-MRS pilot program, 1,200 s was used as the default exposure time to observe the sources with $10\leq G<14.5$ mag. The SNR of the single exposure spectrum of the blue arm would be larger than 5 when $G<14.5$ mag.

%There are not enough sources is too brighter to fill all fibers and will be saturated. 
\begin{figure}
    \centering
    \makebox[\textwidth][c]{\includegraphics[width=\textwidth]{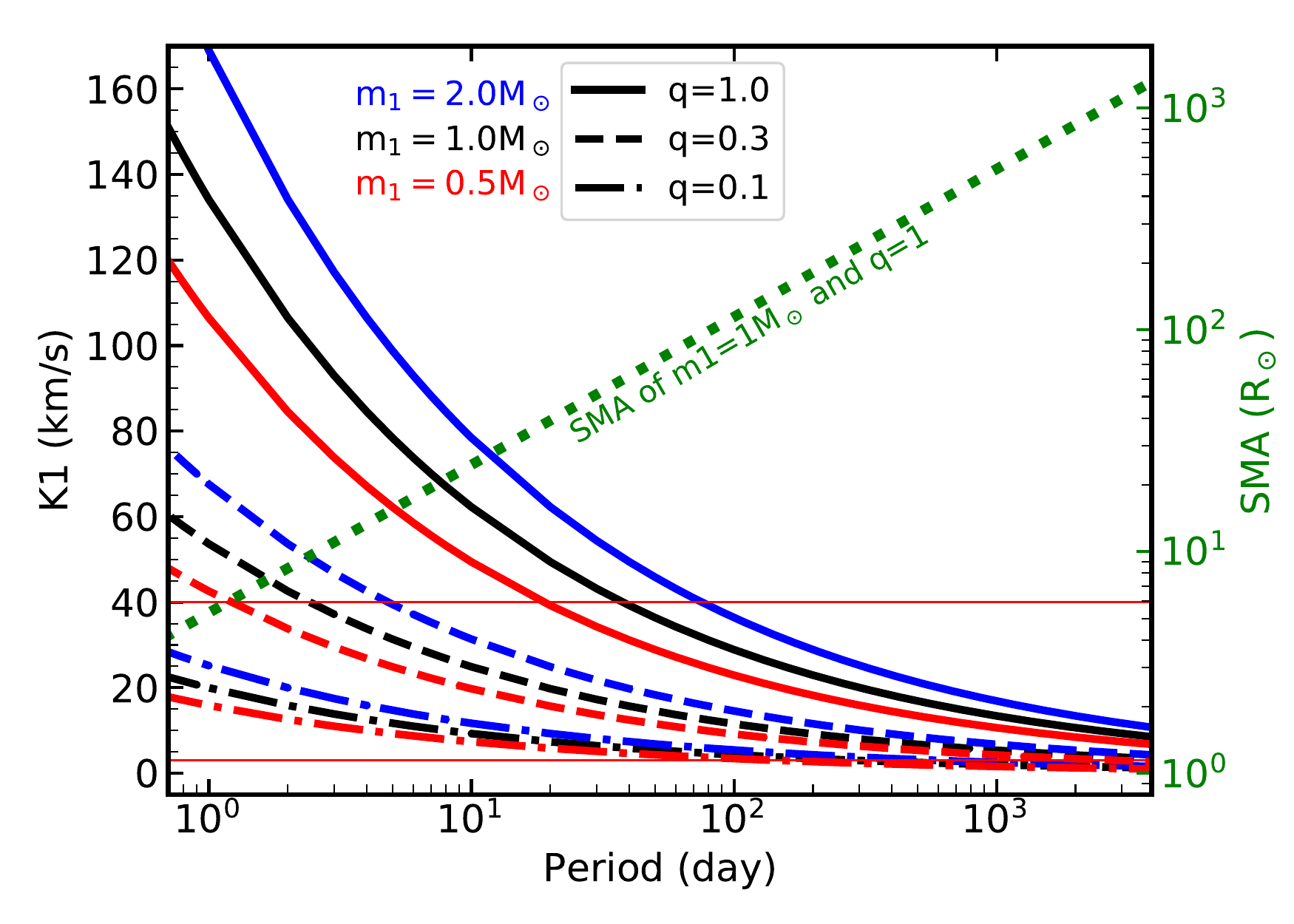}}
    \caption{The semi-amplitude of RV of primary star (semi-major aixs) as a function of period (right ordinate). Here we assume the orbital inclination and eccentricity are $i = 90^\circ$ and $e=0$, respectively. The blue, dark, and red lines stand for different primary mass of $m_1$ = 2, 1, and 0.5 $M_\odot$ respectively. The solid, dashed and dash-dot lines denote mass ratios of $q=$ 1, 0.3 and 0.1, respectively. The top red thin horizontal line denotes the spectrum resolution in RV ($\delta v = c/7500 \sim 40$ \kms). The bottom red horizontal line with rv = 3 \kms. The green line is the SMA of binary with $m_1 = 1~M_{\odot}$ and $q=1$.}
    \label{fig:periodk}
\end{figure}

The binary stars with semi-amplitude of RV ($K > 3\sigma_{\rm rv}$) will be detected with the RV precision $\sigma_{\rm rv} \sim 1$ \kms for late-type stars. Clearly shown in Figure ~\ref{fig:periodk}, LAMOST-MRS can detect the solar-type single-lined spectroscopic binaries (SB1s) with $q>0.1$ and period $p<1000$ days, which would experience interaction in their lifetime. We can also detect the solar-type double-lined spectroscopic binaries (SB2s) with periods $p<50$ days because the RV difference of the two components is larger than 40 (see the top red thin horizontal lines of Figure~\ref{fig:periodk}). The more massive SB2 can be detected with the longer period.

We would like to analyze statistical properties by resolving binary orbital parameters (period, eccentricity, argument of pericenter, phase, and RV amplitude). According to the experiment of the {\it Joker} which is a PYTHON package of deriving orbital parameters more than 10 observations for each star are required to obtain the reliable parameters\cite{Price2017ApJ, Price2018AJ}. In the first year (2018-2019) of LAMOST-MRS, we prepared 25 plates and took exposures as many as possible ($>3$ times) when each plate pass through the observation window of LAMOST. After the observation of the first year (see Table~\ref{tab:planid}), we realized it is difficult to achieve our goal of $>10$ visits for the binaries with $P>10$ days in all 25 plates. Considering the observation Strategies of MSR-K and MRS-T which have the same observational method as the first year of MRS-B (3-8 exposures each visit), the short period binaries could also be found from these two fields. Therefore, we take only three continuous 1200s exposures in each visit in the following time.

\begin{figure}
    \centering
    \makebox[\textwidth][c]{\includegraphics[width=\textwidth]{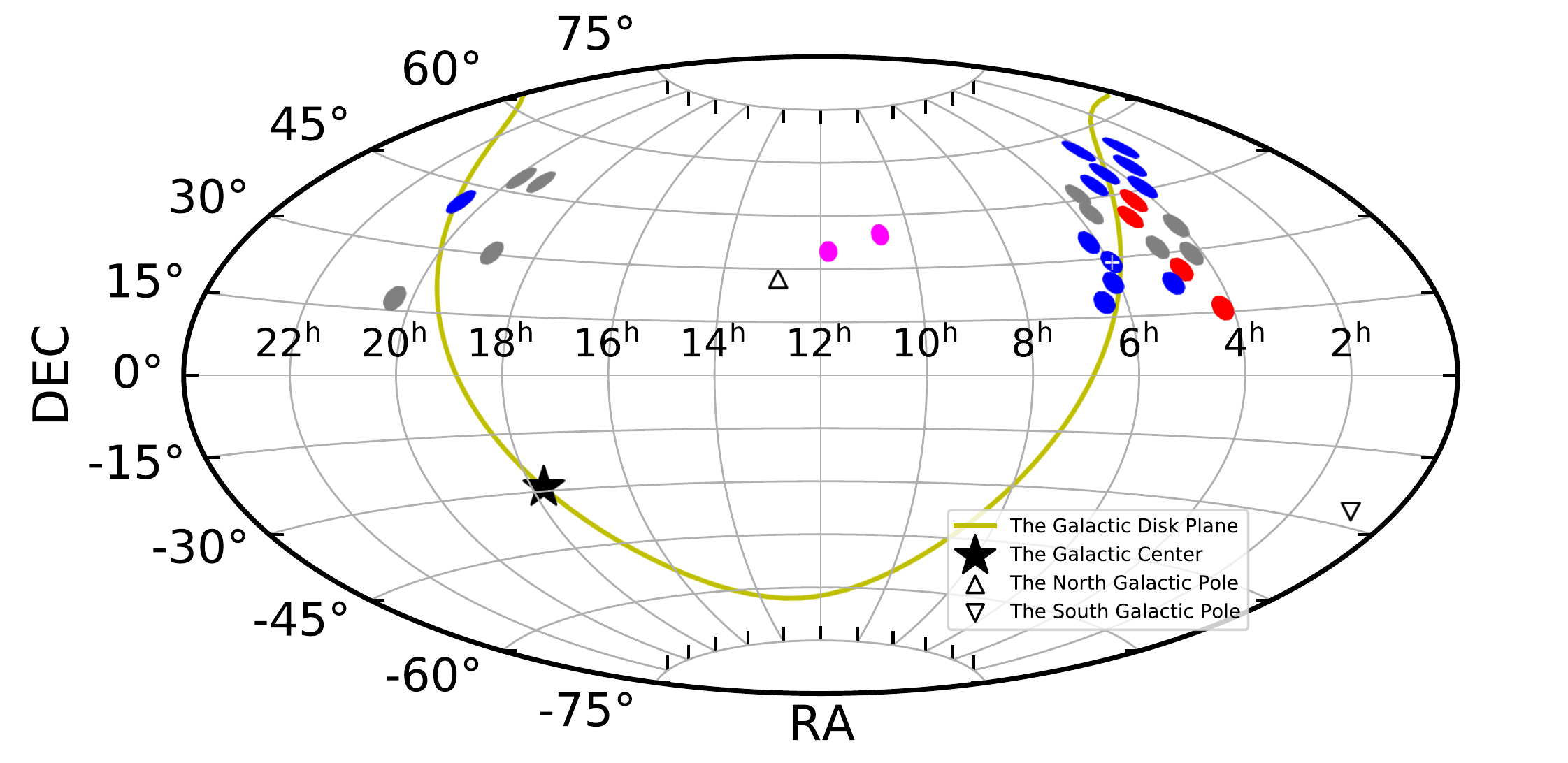}}
    \caption{The Aitoff map shows the footprints of LAMOST-MRS-B project in the celestial coordinate. The 12 blue plates will be observed in the whole 5-year LAMOST-MRS program. The four red plates were only observed at the first year (from 2018-10-05 to 2019-06-08) of the program. The two purple plates were observed from the second year (2019-10-23) of the program and the following years. The nine gray plates have been dropped out without any observation. Note: the blue field labeled white plus is TD055633N285632B01 which has the same center coordinates of HIP28117 which is a pilot program plate of LAMOST-MRS.}
    \label{fig:footprint}
\end{figure}

It is hard to observe each plate with a uniform cadence (or at given dates) on account of the weather condition and the different observation strategies of the six scientific fields. It seems easier to correct the selection bias of orbital periods with random visits. So that the observation dates are randomly selected during the observation season for each plate. In order to detect binaries with period $p > 1$ year, we select two plates TD060648N233818B01 and TD203106N405128B01 to observe as long as five years. 
%We hope to visit the two plate one time per month when they can be observed by LAMOST. However, it seems difficult due to kinds of reasons such as weather condition, . The most important and difficult process of statistic is selection bias correction.

\begin{table}
    \centering
    \begin{tabular}{cc|ccc|ccc|c}
    \hline\hline
    Planid & (ra, dec) & $N_{\rm vist}$&$N_{\rm exp}$&$N^{\rm 1st}_{\rm star}$& $N_{\rm vist}$&$N_{\rm exp}$&$N^{\rm 2nd}_{\rm star}$&$N_{\rm star}$\\
    (B01)    &  ($^\circ$)  &  \multicolumn{3}{c|}{201810-201907}& \multicolumn{3}{c|}{201910-202006}&2018-2020\\ 
\hline
TD030224N542147 & ( 45.60175, 54.36327) &  2 &   7 & 2695 &    5 &  15 & 2892 & 2391\\
TD033106N502853 & ( 52.77737, 50.48149) &  2 &  14 & 2985 &    4 &  12 & 2909 & 2084\\
TD035400N453042 & ( 58.50355, 45.51177) &  1 &   6 & 2942 &    0 &   0 &    0 &    0\\
TD041114N555437 & ( 62.81038, 55.91050) &  2 &  11 & 2755 &    3 &  13 & 2910 & 2257\\
TD042145N500206 & ( 65.43948, 50.03517) &  3 &  16 & 3010 &    1 &   3 & 2807 & 2005\\
TD045606N223435 & ( 74.02765, 22.57656) &  1 &   8 & 2924 &    0 &   0 &    0 &    0\\
TD045704N475243 & ( 74.26917, 47.87884) &  1 &   8 & 3024 &    5 &  16 & 2966 & 2109\\
TD055633N285632 & ( 89.14070, 28.94227) &  2 &  14 & 3025 &    5 &  14 & 2956 & 2279\\
TD060648N233818* & ( 91.70275, 23.63860) &  2 &   8 & 3076 &    3 &   9 & 2951 & 2315\\
TD061044N341537 & ( 92.68602, 34.26042) &  4 &  24 & 3069 &    2 &   6 & 2880 & 2163\\
TD062610N184524 & ( 96.54271, 18.75690) &  6 &  31 & 3213 &    4 &  11 & 2943 & 2180\\
TD203106N405128* & (307.77875,40.85802 )&  3 &   9 & 3189 &    0 &   0 &    0 &    0\\
\hline
TD103949N392416 & (159.95557, 39.40467)&  0 &   0 &    0 &   14 &  48 & 1884 &    0\\
TD114941N345554 & (177.42381, 34.93176)&  0 &   0 &    0 &   15 &  46 & 1713 &    0\\
\hline
TD041055N155647$^{\rm x}$ & ( 62.72986, 15.94666)&  1 &   4 & 2141 &    | &   | &    | &    |\\
TD042649N425434$^{\rm x}$ & ( 66.70605, 42.90949)&  1 &   6 & 3000 &    | &   | &    | &    |\\
TD043724N254338$^{\rm x}$ & ( 69.35248, 25.72738)&  2 &  11 & 2909 &    0 &   0 &    0 &    0\\
TD045227N391702$^{\rm x}$ & ( 73.11599, 39.28412)&  1 &   6 & 2972 &    | &   | &    | &    |\\
\hline
TD040433N355514$^{\rm x}$ & ( 61.14068, 35.92056)&  | &   | &    | &    | &   | &    | &    |\\
TD041324N290722$^{\rm x}$ & ( 63.35187, 29.12292)&  | &   | &    | &    | &   | &    | &    |\\
TD044912N312614$^{\rm x}$ & ( 72.30353, 31.43737)&  | &   | &    | &    | &   | &    | &    |\\
TD053403N461112$^{\rm x}$ & ( 83.51411, 46.18688)&  | &   | &    | &    | &   | &    | &    |\\
TD053908N412130$^{\rm x}$ & ( 84.78624, 41.35856)&  | &   | &    | &    | &   | &    | &    |\\
TD185549N301843$^{\rm x}$ & (283.95518, 30.31206)&  | &   | &    | &    | &   | &    | &    |\\
TD191631N482124$^{\rm x}$ & (289.12983, 48.35674)&  | &   | &    | &    | &   | &    | &    |\\
TD195013N482329$^{\rm x}$ & (297.55489, 48.39158)&  | &   | &    | &    | &   | &    | &    |\\
TD202021N174734$^{\rm x}$ & (305.08919, 17.79293)&  | &   | &    | &    | &   | &    | &    |\\
\hline\hline
    \end{tabular}
    \caption{Plates of MRS-B field, where $N_{\rm vist}$, $N_{\rm exp}$, $N_{\rm star}$ are the number of visits (observation dates), exposure times, and observation stars, respectively. $N^{\rm 1st}_{\rm star}$ and $N^{\rm 2nd}_{\rm star}$ are the number of stars observed at the first and second year of regular surveys, respectively. The plates labeled by symbols of $^{\rm x}$ will not observe after October 2019 in the field of MRS-B. * means that the two plates will be observed all five year of the program. TD103949N392416 and TD114941N345554 were inputted after October 2019. The last column stand for the number of common stars observed in the both years. The observation priorities of stars were modified since the second years so that $N_{\rm star}$ is smaller than $N^{\rm 1st}_{\rm star}$ or smaller than $N^{\rm 2nd}_{\rm star}$. Note: TD043724N254338 will be observed as plate of the MRS-S field; in these table, we count all stars including the source both \snrb{$=0$} and \snrr{$=0$}.}
    \label{tab:planid}
\end{table}

\section{Target Selection}
Most massive stars are located in the Galactic disk. The companions of SB1s with early-type MS primaries may not necessarily be low-mass A-K type stars. They may contain 1-3 $M_\odot$ stellar remnants such as WDs, neutron stars, or even BHs\cite{Wolff1978ApJ, Garmany1980ApJ, Shaoyong2020ApJ}. In addition, the unresolved solar-type SB1s might contain compact companions as well. Most of the compact companions are WDs\cite{Hurley2002MNRAS, Belczynski2008ApJS} accompanied by UV excess. It is useful to select solar-type + WD binary by using UV sources catalogs of the Galaxy Evolution Explorer (GALEX)\cite{Bianchi2017ApJS}. However, the Galactic disk was not observed by the FUV-detector of GALEX (see Figure 1 of Bianchi et al. 2017). To pick out these rare exotic objects as many as possible, and sufficiently utilize the 3000 fibers of LAMOST, we initially selected 25 plates around the disk (see blue + red + gray plates in Figure~\ref{fig:footprint}).

\begin{figure}
    \centering
    \makebox[\textwidth][c]{\includegraphics[width=0.7\textwidth]{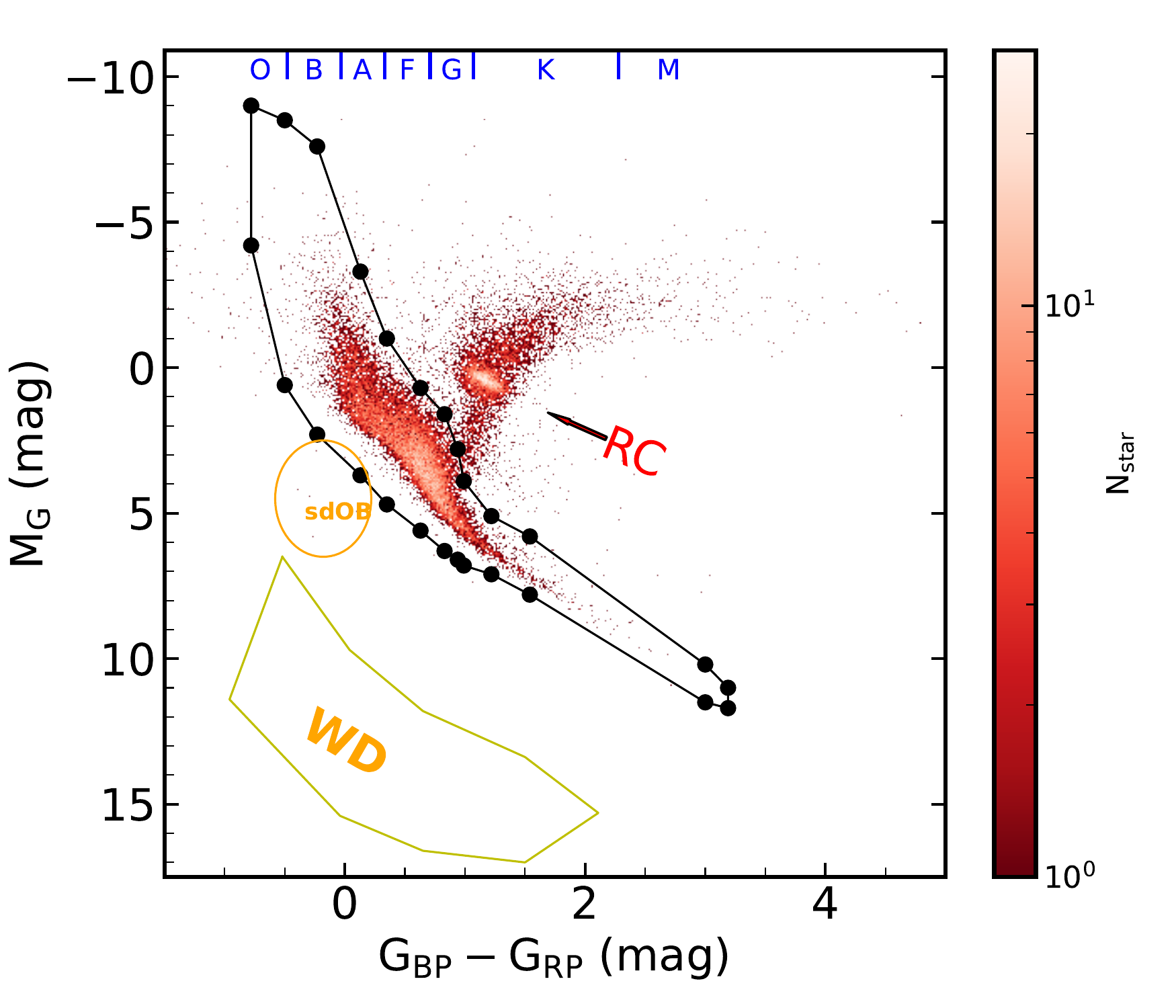}}
    \caption{{\it Gaia} Color-Magnitude diagram of source of MRS-B field in DR8. The color and $G$ magnitude have been corrected by using \textbf{dustmaps}\cite{Green2019ApJ}. The black polygon indicate the location of MS stars, and its fixed points are \bprp{= (-0.78, -0.5, -0.23, 0.13, 0.35, 0.63, 0.83, 0.94, 0.99, 1.22, 1.54, 3.0, 3.19, 3.19, 3.0, 1.54, 1.22, 0.99, 0.94, 0.83, 0.63, 0.35, 0.13, -0.23, -0.5, -0.78)}, $\MG$ = (-9, -8.5, -7.6, -3.3, -1.0, 0.7, 1.6, 2.8, 3.9, 5.1, 5.8, 10.2, 11.0, 11.7, 11.5, 7.8, 7.1, 6.8, 6.6, 6.3, 5.6, 4.7, 3.7, 2.3, 0.6, -4.2). The orange curves denote the locations of sdOB and WD. RC is red clump star.} %http://dr6.lamost.org/v2/doc/lr-data-production-description
    \label{fig:cmd}
\end{figure}

Numbers of massive stars (O- and B-type), sdOBs, and WDs are selected with help of {\it Gaia} color-magnitude diagram (CMD) as shown in Figure~\ref{fig:cmd}. All those samples are set with highest observation priority in the input catalog. The massive stars were picked out with $G_{\rm BP} - G_{\rm RP} < 0$; the sdOB and WD candidates are selected by the sources under the MS stars in the CMD. Here, we didn't correct the redden of \bprp{} and extinction of $\MG$. To obtain the spectra with good qualities, for the rest of the stars with $10\leq G<14.5$, the observation priorities followed the rule: the stars that are in the catalog of GALEX and brighter would have higher observation priority (details see the second column of Table~\ref{tab:priority}). %It is independent of studying the statistic properties of binaries.

\begin{table}
    \centering
    \begin{tabular}{c|c|c|c}
    \hline\hline
    Priority & 201810-201907    & (LGL plates) 201910-                          & (HGL plates) 201910-\\
\hline  
1 & ${\rm G_{OBWD}}$            &  $SE_{18-19}$                                 &   LFeH \& ($10\leq G<14.5$) \\
2 & GALEX \& ($10\leq G<13$)    &  !$SE_{18-19}$ \& ($10\leq G<14.5$)           &   $SE$ \& ($10\leq G<14.5$) \\
3 & !GALEX \& ($10\leq G<12$)   &  $S_{18-19}$ \& LFeH                          &   ${\rm G_{OBWD}}$ \\
4 & !GALEX \& ($12\leq G<13$)   &  rest LFeH \& ($10\leq G<14.5$)      &   GALEX \& ($10\leq G<13$)    \\
5 & GALEX \& ($13\leq G<14$)    &  rest $S_{18-19}$                             &   !GALEX \& ($10\leq G<12$)   \\
6 & GALEX \& ($14\leq G<14.5$)  &  rest ${\rm G_{OBWD}}$                        &   !GALEX \& ($12\leq G<13$)   \\
7 & !GALEX \& ($13\leq G<14$)   &  rest GALEX \& ($10\leq G<13$)                &   GALEX \& ($13\leq G<14$)    \\
8 & !GALEX \& ($14\leq G<14.5$) &  rest !GALEX \& ($10\leq G<12$)               &   GALEX \& ($14\leq G<14.5$)  \\
9 &                             &  rest !GALEX \& ($12\leq G<13$)               &   !GALEX \& ($13\leq G<14$)   \\
10&                             &  rest GALEX \& ($13\leq G<14$)                &   !GALEX \& ($14\leq G<14.5$) \\
11&                             &  rest GALEX \& ($14\leq G<14.5$)              &                               \\
12&                             &  rest !GALEX \& ($13\leq G<14$)               &                               \\
13&                             &  rest !GALEX \& ($14\leq G<14.5$)             &                               \\
\hline\hline
    \end{tabular}
    \caption{The priority of input source. The second column are priority of input sources in the first year (201810-201907). The third and last columns are for plates located close to the Galactic disk and at the high Galactic latitude, respectively(see, 12 blue and two purple plates of Figure~\ref{fig:footprint}).  ${\rm G_{OBWD}}$ stands for O-, B-, WD and sdOB selected from {\it Gaia} CMD ($10\leq G<14.5$, without correcting color and extinction). ! indicate the source without observation. $S_{18-19}$ denotes star observed in the first year of LAMOST-MRS ;  $SE$ means the stars in the our exotic stars catalog (see Table~\ref{tab:exotic}); $SE_{18-19}$ stand for the exotic star observed in the first year of LAMOST-MRS. LFeH stands for the low-metallicity candidate selected by line indices of ${\rm H\alpha}$ and CaII K (see Figure~\ref{fig:ewhacak}). }
    \label{tab:priority}
\end{table}

After the first year observation of LAMOST-MRS, we found that it is difficult to obtain $>10$ visits for all the 25 plates of which only 16 plates were observed with only 34 visits. Most TD plates were crowd around the Galactic disk (see Figure 2 of Liu et al 2020\cite{Liuchao2020arxiv}), so they are prone to take place ``traffic jams''. Therefore, we droped out 13 plates (see Table~\ref{tab:planid}), where TD043724N254338 would be observed in the field of MRS-S, and added two new plates located at the high Galactic latitude (TD103949N392416, TD114941N345554, as shown in the purple plates of Figure~\ref{fig:footprint}). We also modified the observation priorities of input catalogs on account of studying the variable stars and statistic properties of low-metallicity stars.

The periods and some physic parameters of exotic stars can be figured out by the spectra of TD observation. According to the {\it Understanding Variable Stars}\cite{Percy2007uvs} and other catalogs, we compiled a catalog (hereafter, we name it exotic star catalog) with 112,554 exotic stars which include Be stars, Cataclysmic Variable stars, Classical Cepheid stars, X-ray binaries, symbiotic stars and so on (see Table~\ref{tab:exotic}, the catalog is available in Science Data bank).

\begin{table}
    \centering
    \begin{tabular}{c|c|c|c}
    \hline\hline
    Name & Brief description & $N_{\rm star}$ &References\\
\hline
${\rm Be_{HIPP}}$   &Be star in the HIPPARCOS                                   & 0&\cite{1998A&A...335..565H}\\
CV                  &Cataclysmic Variables                                      & 0&\cite{2001PASP..113..764D}\\
Cepheid             &Classical Cepheid stars                                    & 1&\cite{2003IBVS.5394....1S,2017ARep...61...80S}\\
${\rm EB_{Algol}}$  &Eclipsing binaries with Algol-type light curve             & 2&\cite{2004A&A...417..263B,2018ApJS..238....4P}\\
${\rm EB_{Xray}}$   &Eclipsing binaries with Xray                               & 2&\cite{2000A&AS..147...25L, 2006A&A...455.1165L,2007A&A...469..807L}\\
GCVS5               &the General catalogue of variable stars: Version GCVS 5.1  &149&\cite{2017ARep...61...80S}\\
$\gamma$Dor         &$\gamma$ Doradus variable stars                            & 0&\cite{1990A&AS...86..107H, 1995IBVS.4216....1H,1995IBVS.4195....1K}\\
$\gamma$Ray         &$\gamma$ Ray Burst Pulses                                  & 0&\cite{2014ApJS..211...12G,2014ApJS..211...13V,2016ApJS..223...28N, 2018ApJ...855..101H}\\
Herbig              &Herbig Ae Be stars                                         & 6&\cite{ 1994A&AS..104..315T,Herbst1999AJ, 2018A&A...620A.128V}\\
LumStar             &Photometry and Spectroscopy for Luminous Stars             &168&\cite{2005AJ....130.1652R}\\
OB                  &O- and B-type stars                                            &1851&\cite{2017A&A...598A..84A,Liuzhicun2019ApJS}\\
RCB                 &The R Coronae Borealis stars                               & 0&\cite{1996PASP..108..225C,2018AJ....156..148M,2019MNRAS.482.4174F}\\
RRLyrae             &RR Lyrae stars                                             & 1&\cite{2005A&A...442..381M}\\
RotateVB            &Rotational variable stars                                  & 0&\cite{2005AJ....129.1993M,2008MNRAS.389.1722E,2009A&A...498..961R,2013A&A...552A..78Z,2018A&A...616A.108B}\\
SB                  &Spectroscopy binaries                                      & 0&\cite{2016AJ....151....3G}\\
SySt                &Symbiotic stars                                            & 0&\cite{Lijiao2015RAA,Akras2019ApJS}\\
TTauri              &T Tauri stars                                              & 6&\cite{2015A&A...580A..26G,2015A&A...580A..26G}\\
UVCeti              &the UV Cet-type flare stars                                & 0&\cite{1999A&AS..139..555G}\\
WD+FKG              &WD+FKG-type binaries                                       & 6&\cite{Rebassa-Mansergas2017MNRAS}\\
WR                  &Wolf Rayet stars                                           & 1&\cite{1968MNRAS.138..109S,2001NewAR..45..135V,2015MNRAS.447.2322R}\\
$\delta$ Scuti      &$\delta$ Scuti stars                                       & 5&\cite{2013AJ....145..132C}\\
\hline\hline
    \end{tabular}
    \caption{Catalogue name of exotic stars (their coordinates can be found in Science Data bank, Catalog 2). $N_{star}$ denotes the number of stars in the MRS-B field of DR8 which accumulate all LAMOST-MRS spectra observed from 2017-09-19 to 2020-05-27.}
    \label{tab:exotic}
\end{table}

\begin{figure}
    \centering
    \makebox[\textwidth][c]{\includegraphics[width=0.7\textwidth]{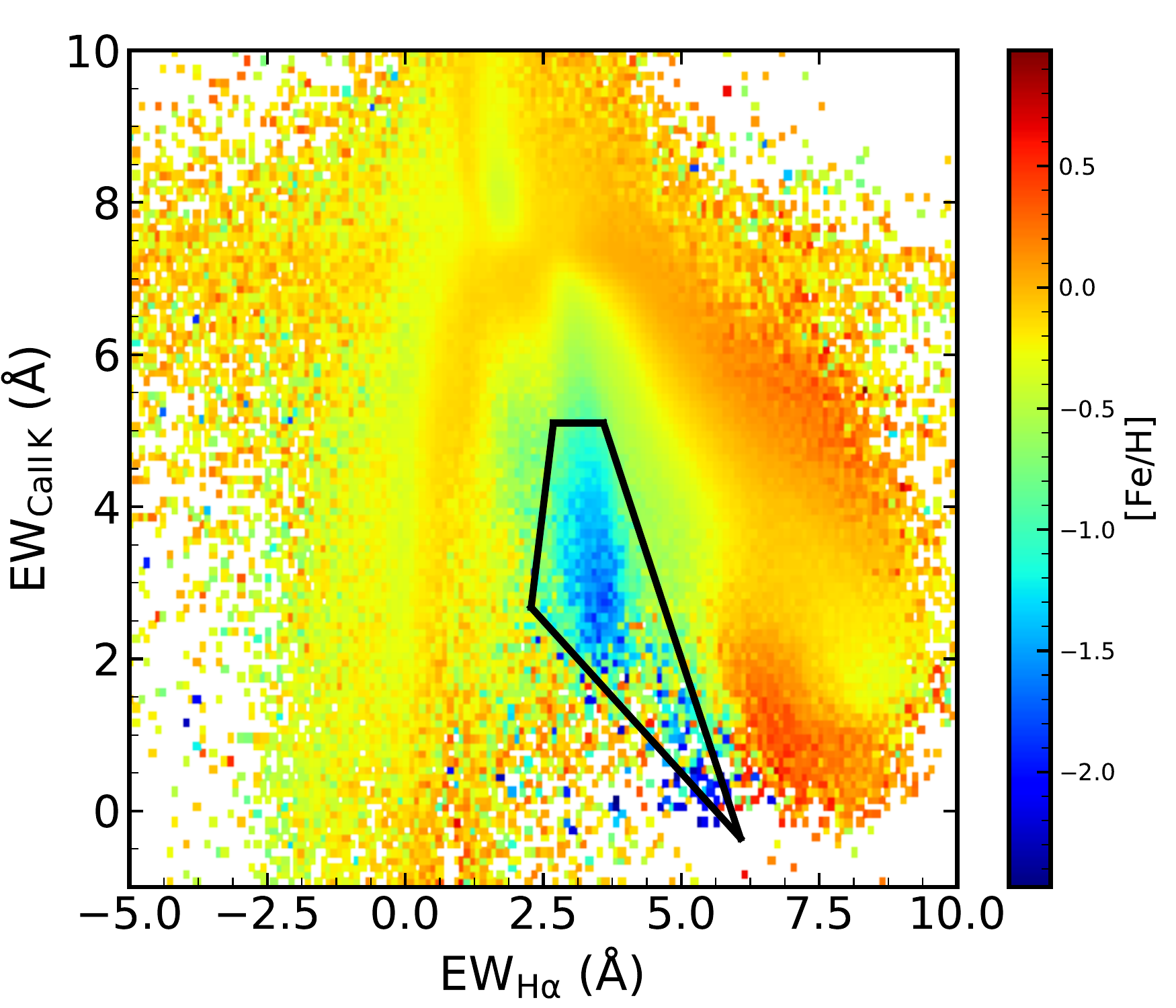}}
    \caption{Metallicity ([Fe/H]) as a function of the equivalent width of ${\rm H\alpha}$ and CaII K. The line indices are calculated from the spectra in the DR6 of LAMOST-LRS. The iron abundance is derived by the LAMOST pipeline. 
    The index band-pass of $\rm H\alpha$ and CaII K are 6548-6578 and 3927.7-3939.7 \AA\cite{Liu2015RAA}. The four fixed point of the black polygon are (2.683, 5.1), (3.59, 5.1), (6.071, -0.357), (2.279, 2.676).} %%http://dr6.lamost.org/v2/doc/lr-data-production-descript%ion
    \label{fig:ewhacak}
\end{figure}

Although close binary fraction decreases with metallicity proved by the recent research \cite{Moe2019ApJ,  Mazzola2020MNRAS}, it is still unclear whether the relation is universal for all primary stars. However, the low-metallicity (${\rm [Fe/H] < -1}$) stars are rare. We pick out the low-metallicity candidates from LAMOST-LRS spectra of DR6 by using the diagram of line indices of ${\rm H\alpha}$ and CaII K. The index band-pass of the line indices can be found in Table 2 of Liu et al. (2015) \cite{Liu2015RAA}. From Figure~\ref{fig:ewhacak}, we can see that low-metallicity stars are concentrated at a specific region in the diagram of line indices of ${\rm H\alpha}$ and CaII K. We roughly selected the sources located in the black polygon as our low-metallicity candidates.

Since the identified exotic stars and low-metallicity stars are rare, we increased their priorities in the following observations. Then we gave the observed sources high priorities to guarantee that we would observe plenty of stars with $>10$ visits (see Table~\ref{tab:priority}). 

% to guarantee the stars have as many as possible of visits, we

\section{Parameter determination and analysis}

Raw CCD data were reduced by the LAMOST 2D pipeline. The procedures of extracting 1D spectra from 2D images include dark and bias subtraction, spectral extraction, sky subtraction and wavelength calibration. The procedures are the same as those of the 2D pipeline used by the LAMOST-LRS\cite{Luo2015RAA}, except that the wavelength calibration is based on Th-Ar or Sc lamps without stacking of subexposures\cite{Zhangbo2021ApJS}. The SNRs of all pixels were calculated, and the medians were taken as the final values for blue (\snrb{}) and red (\snrr{}) arms of each spectrum. The Eighth year survey data has been released, which includes LAMOST-MRS spectra observed from September 2018 to Jun 2020 (DR8-MRS\footnote{http://www.lamost.org/dr8/v1.1}). There are 53,360 sources observed in MRS-B field, but \snrb{} and \snrr{} of some sources are both zeros due to the bad fibers or weather conditions. We picked out 731,752 spectra with \snrb{$>5$} or \snrr{$>5$} that belong to 49,129 sources of MRS-B. Crossmatching the sources with DR8-MRS by the sky separation of 5$"$, we obtain 892,233 spectra with \snrb{$>5$} or \snrr{$>5$}, because some of the sources are observed by the other sub-surveys of LAMOST-MRS (such as NT survey) as well. We get 28,828 and 3,375 sources that are visited $\geq 2$ and $\geq 10$ times, which have at least a \snrb{$\geq 10$} spectrum in each visit.

\subsection{Atmospheric parameters}

The stellar parameters are from the LAMOST pipeline, i.e., LASP, which adopts the ELODIE library as templates. The templates are interpolated with a polynomial spectral model and then downgraded to LAMOST spectral resolution automatically while matching to observations \cite{Wuyue2011RAA}. In DR8, the published LAMOST MRS stellar parameters are derived from the stacked spectra from multi-epoch spectra of the blue arm obtained in one night.
Zong et al. (2020) found the uncertainties of $T_{\rm eff}$, $\log g$ and [Fe/H] are 100 K, 0.15 dex and 0.09 dex, respectively\cite{Zongweikai2020ApJS}, by comparing the common sources observed by {\it Gaia}\cite{Andrae2018A&A} and the Apache Point Observatory Galactic Evolution Experiment (APOGEE)\cite{Majewski2017AJ, Ahumada2020ApJS} in the MRS-K field. There are 36,516 sources in the MRS-B field that have atmospheric parameters derived by LAMOST pipeline (LASP).

\begin{figure}[!ht]
    \centering
    \subfloat{\includegraphics[width=0.45\textwidth]{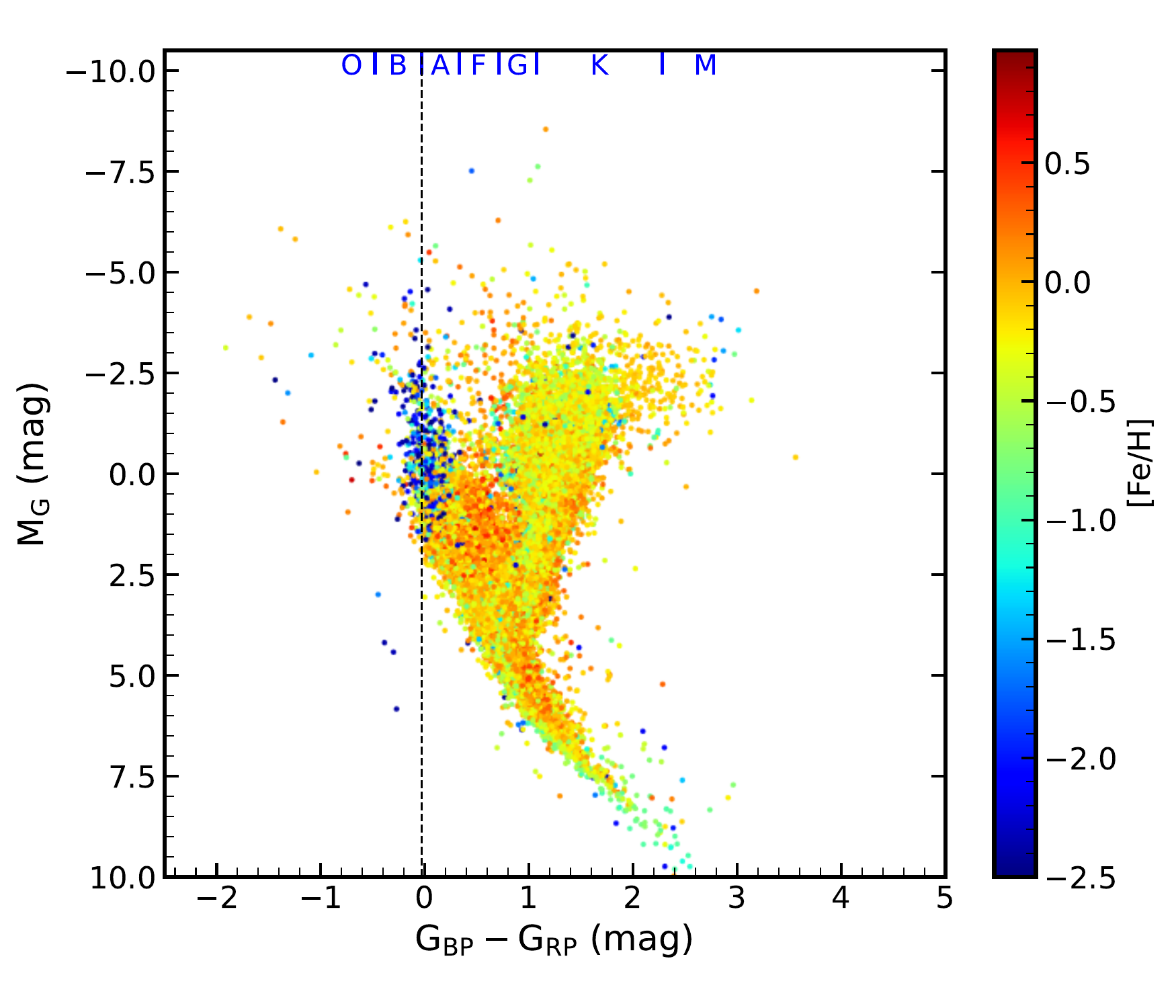}}(a) 
    \subfloat{\includegraphics[width=0.45\textwidth]{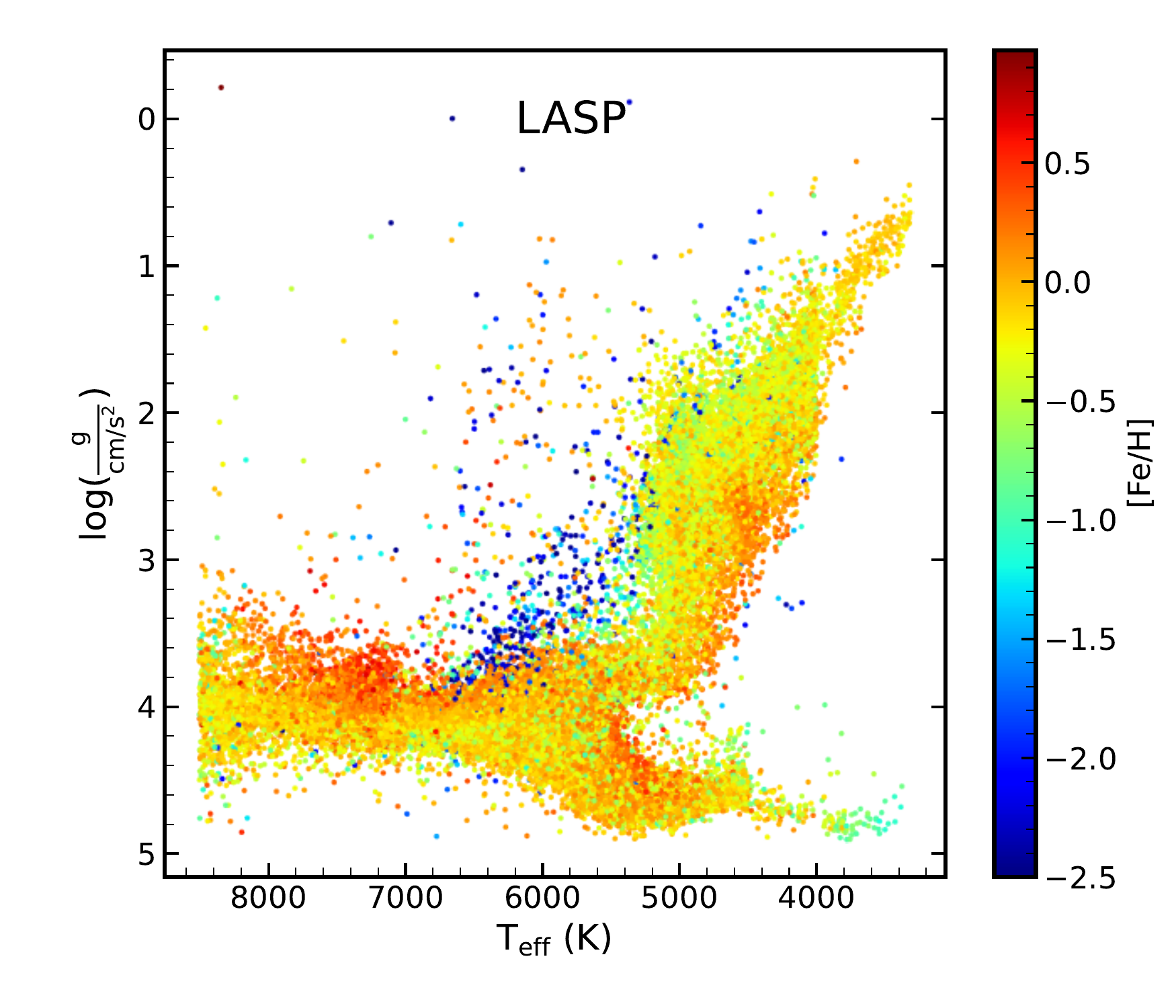}}(b) 
    \subfloat{\includegraphics[width=0.45\textwidth]{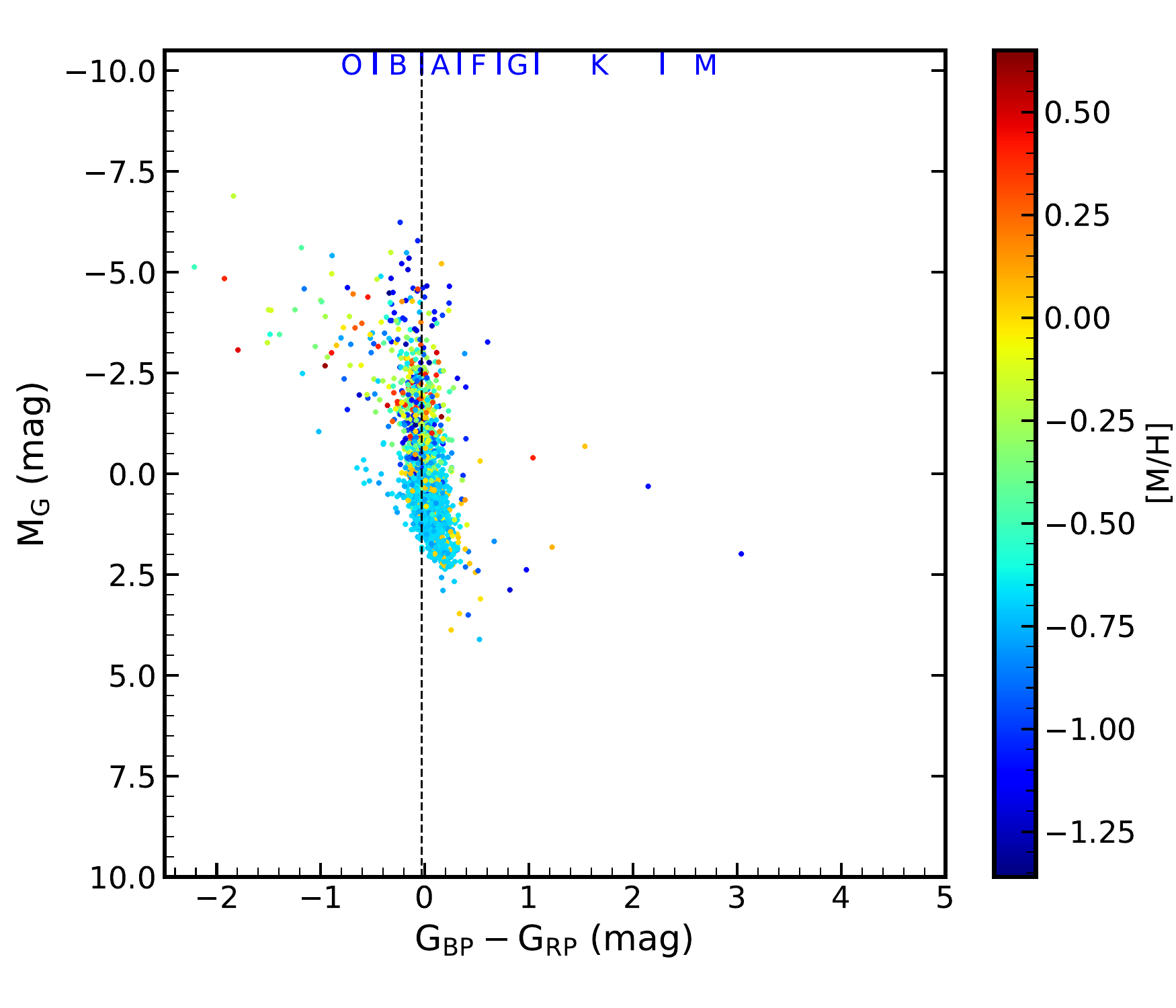}}(c)
    \subfloat{\includegraphics[width=0.45\textwidth]{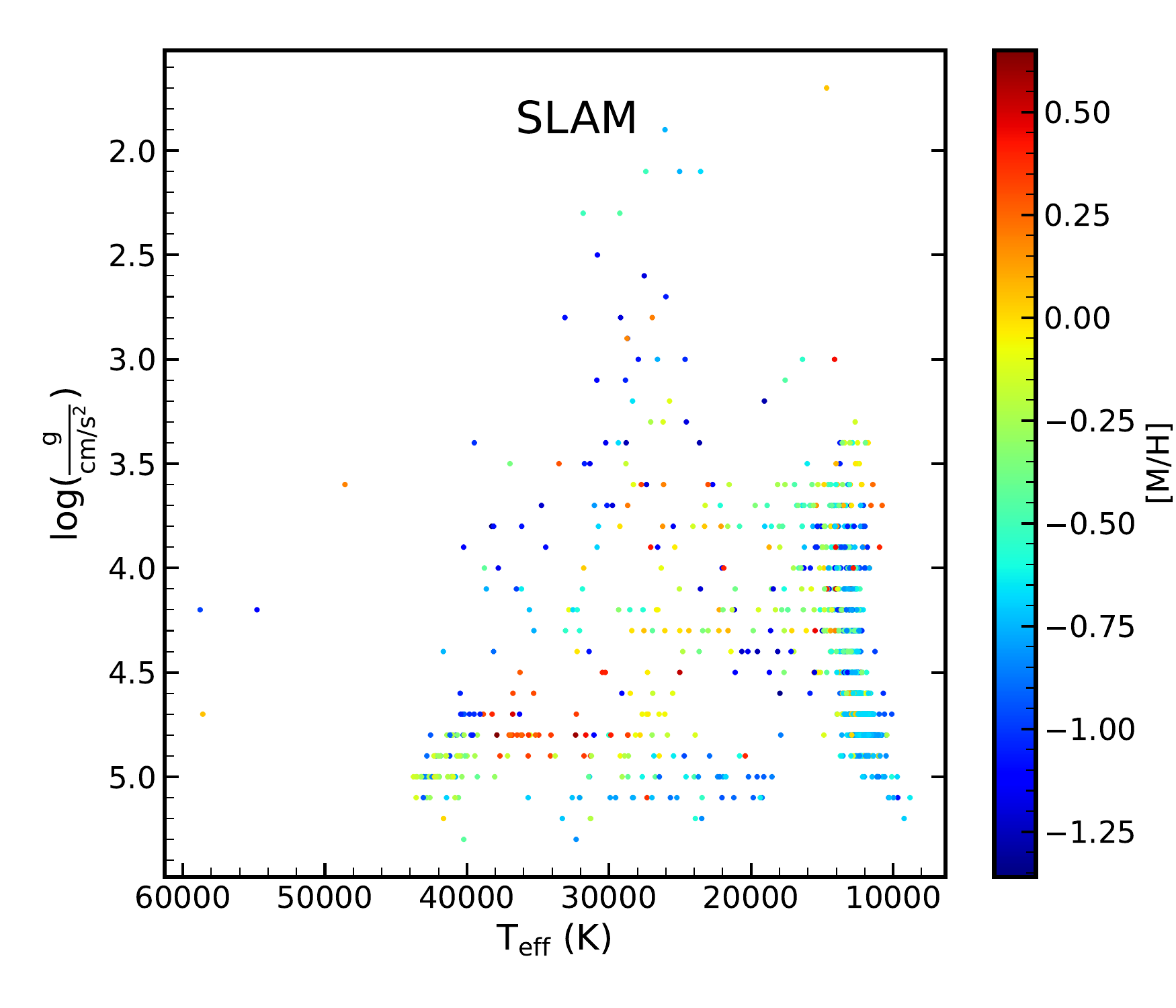}}(d)
    \caption{The CMD and Kiel diagram colored by [Fe/H] or [M/H]. The upper rows show the stars with atmosphere parameters derived by the updated LASP pipeline, the  median values are use for the source with multiple spectra. The bottom rows show the early-type stars with parameters measured with the best SNR spectra by using SLAM. Since the uncertainties is about 0.29 dex, only one decimal place is saved, so that the $\log g$ in the panel (b) is waffle-like. Note that: [M/H] is the abundance of elements present in an object that are heavier than hydrogen and helium; [Fe/H] is the iron abundance, we use [Fe/H] to stand for metallicity abundance if there is no specific statement. {\it Caution}: there are two stars (obsid: 727414208, 681809175) with $T_{\rm eff} > 5,5000$ K, which are misled by the features of SB2.} %http://dr6.lamost.org/v2/doc/lr-data-production-description
    \label{fig:cmd_kiel}
\end{figure}

A few contamination of early-type spectra can be found, as shown the blue dots in panels (a) and (b) of Figure~\ref{fig:cmd_kiel}. Some early-type stars with corrected \bprp{$< 0.1$} would be mistaken as metal-poor stars with temperature $T_{\rm eff} \sim 6000$ K. Because the LASP could only provide parameters for stars with spectral types of late A-, F-, G-, and K-type due to the limitation of templates library. The early-type stars, especially O- and B-type stars barely have metal lines in the blue arm spectra which thus make them look like poor metal-poor stars.

Guo et al. (2021) picked out early-type stars from DR7 and selected the best SNR epoch spectra to measure the atmospheric parameters by using the Stellar LAbel Machine (SLAM) \cite{Guoyanjun2021ApJS, Zhangbo2020ApJS}. We find 1,718 early-type stars of MRS-B field are in their Table 2 that atmospheric parameters was derived by using LAMOST-MRS spectra. According to their analysis, the uncertainties of $T_{\rm eff}$, $\log g$ are 2,185 K and 0.29 dex; the metallicity abundance [M/H] can't be constrained very well with LAMOST-MRS spectra (details see the bottom-left panel of their Figure 9). There are two stars (see Figure~\ref{fig:cmd_kiel} d, obsid: 727414208, 681809175) with $T_{\rm eff} > 55, 000$ K which are out of the training grid of SLAM (10,000 - 55,000 K). We find that their spectra has features of SB2 checked by eyes, which misled the measurements of effective temperature.

\subsection{Stellar Type}

There is a gap ($8500<T_{\rm eff} < 10000$ K) between the $T_{\rm eff}$ derived by the LASP and the SLAM, and the atmospheric parameters are not derived for all of the sources in the MRS-B filed. For the sake of simplicity and uniformity, we choose {\it Gaia} CMD to coarsely identify which types the sources belong to (see Figure~\ref{fig:cmd}). We correct the color and magnitude induced by interstellar dust. The color excess $E(B-V)$ is obtained by using the median value of the Bayestar 3D dust maps of \textbf{dustmaps} \cite{Green2019ApJ} with the galactic coordinates, parallax of {\it Gaia} EDR3\cite{Gaiaedr32021A&A}, and the distance is simply derived by dividing the parallax ($d=1/\varpi$). Then the color excess $E(G_{\rm BP}-G_{\rm RP})$ and extinction $A_G$ are calculated with the following equations,
\begin{equation}
    E(G_{\rm BP}-G_{\rm RP}) = (1.002 - 0.589)\frac{E(B-V)}{0.317},
\end{equation}

\begin{equation}
    A_G = 0.789\frac{E(B-V)}{0.317},
\end{equation}
where the coefficient are relative extinction values of Table 3 of Wang et al. (2019) \cite{Wangshu2019ApJ}.

\begin{figure}
    \centering
    \makebox[\textwidth][c]{\includegraphics[width=\textwidth]{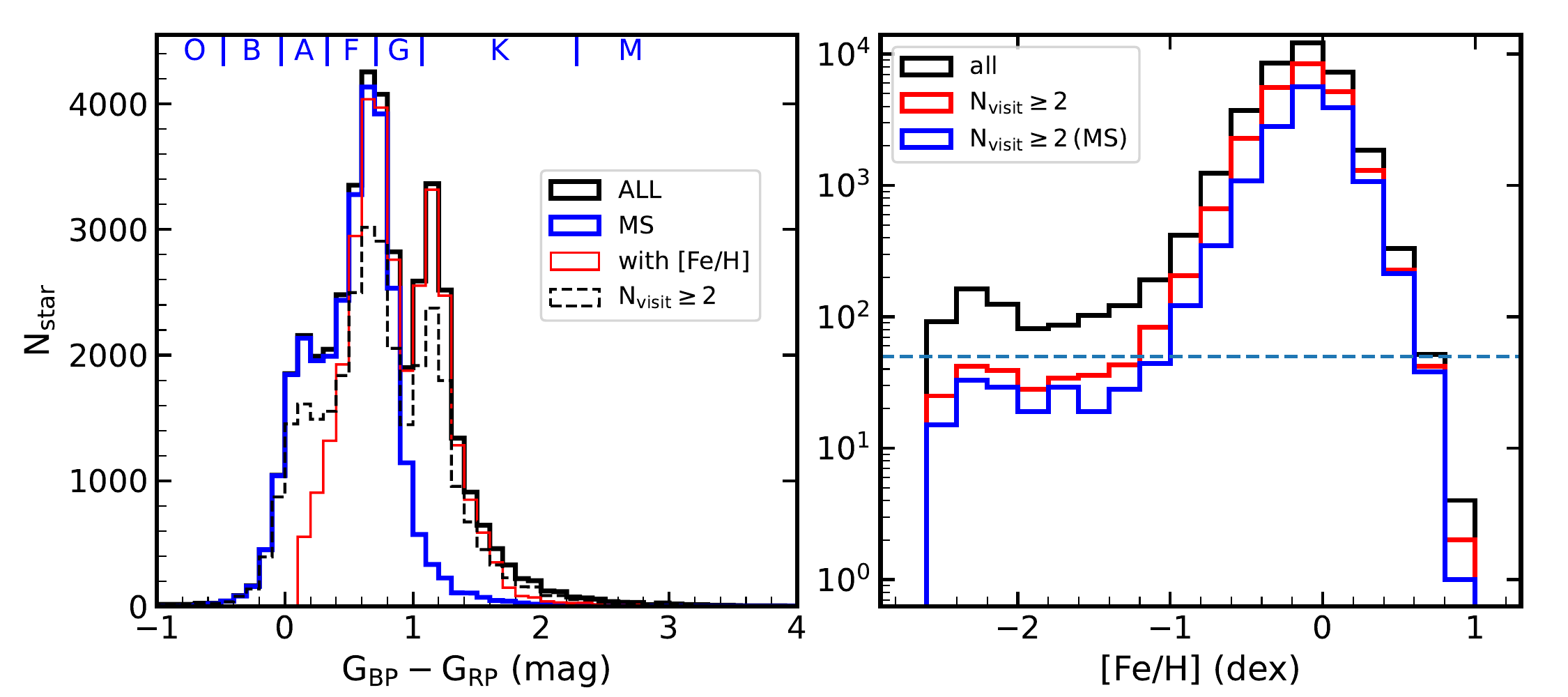}}
    \caption{The histogram of stars which have at least a spectrum of \snrb{$ \geq 10$} in each visit. MS stands for main sequence star. \bprp{} is corrected by using \textbf{dustmaps}. The ordinate axis of the right panel is in a logarithmic scale to make the small numbers visible at poor and rich metallicity bins.} %http://dr6.lamost.org/v2/doc/lr-data-production-description
    \label{fig:colorfeh}
\end{figure}

There are 49,078 stars with available color index \bprp{} and parallaxes. We can see that majority of the stars are MS and Red Clump stars (RCs), and F-type dwarf stars are maximum which contain 7927 sources with \nvisitt{} and each visit has at least a spectrum of \snrb{$\geq 10$} (hereafter, \nvisitt{} denotes the number of stars that have been visited more than twice and have at least spectrum of \snrb{$\geq 10$} in each visit without specification explanation). More than 1,500 sources are observed with \nvisitt{} for each of B-, A-, F-, and G-type MSs (for details see Table~\ref{tab:nstar_type}). We coarsely select MS star by using the black polygon of Figure~\ref{fig:cmd}, since it does not influence statistical result so much. Giant stars can be observed in a larger volume than the volume of dwarf stars because LAMOST-MRS-B only observe the stars with apparent magnitude from 10 to 15 mag. So that most G- and K-type stars are giants and located in the region of the RC star (see Figure~\ref{fig:cmd}), as shown in the left panel of Figure~\ref{fig:colorfeh}. We can see that K- and M-type dwarfs are less; G-type dwarfs are even less numerous than F-type (see left panel of Figure~\ref{fig:colorfeh} and Table~\ref{tab:nstar_type}) because we prefer to observe brighter stars to obtain higher SNRs.

Although the highest observation priorities are given to the exotic stars, only few of them has been observed. As the orange curves shown in Figure~\ref{fig:cmd}, we select WD and sdOB. Only one WD (J041010.32+180223.8) is observed and cannot even see in the 2D histogram of the CMD. It is faint with $G = 11.42$ mag, \snrb{=7.14} and \snrr{=5.24}. Nine sdOB are observed one of which has been visited 10 times. WD and sdOB are too faint or rare to be observed. Eliminating O- and B-type stars, 347 stars of the exotic star catalog are (see Table~\ref{tab:exotic}) in the MRS-B of DR8, which include a Wolf Rayet star (HD 195177) and two High Mass X-ray Binaries (IGR J06074+2205 and HD 249179). 

% even though we gave them the highest observation priority
%  In addition, the smaller stars will evolve longer time to take place mass transfer. Some of them even cannot experience interaction. We mainly consider about BPS induced from the binary interaction. In this aspect, the smaller number of M-type stars will not influence the results of BPS very much.

The low [Fe/H] candidates also have high observation priorities. Excluding the contamination of O- and B-type stars (\bprp{$<0.1$}), we have only 276 stars with ${\rm [Fe/H] < -1}$ dex. It is hard to obtain the statistic property of binary orbital parameters for the metal-poor stars. As shown in the right panel of Figure~\ref{fig:colorfeh}, most stars are distributed in the interval of $-0.6 <{\rm [Fe/H]} < 0.4$ dex. Thousands of them are observed more than twice. There are two stars with ${\rm [Fe/H] > 0.8}$. {\it Note}: the derived [Fe/H] might be wrong for the the metal-richest and metal-poorest.

We compile a catalog that includes RVs, stellar atmospheric parameters derived by LASP and SLAM, information of exotic stars, and a few columns of LAMOST and {\it Gaia} such as {\it obsid}, {\it mjd}, {\it parallax} and {\it source\_id}, for the 892,233 spectra (the catalog is available in \href{http://www.doi.org/10.57760/sciencedb.j00113.00035}{Science Data bank}).

%From the Table~\ref{tab:exotic} and Figure~\ref{fig:cmd}, we can see that only few exotic stars, sdOB are observed even improving their priority.  34272 source are left. There are

%\begin{table}
%    \centering
%    \begin{tabular}{cccccccc}
%    \hline\hline
%    TYPE  & O     & B           & A          & F         & G          & K          & M\\
%    \bprp & $<-0.3$ & (-0.3, 0.1) &(0.1, 0.5)  &(0.5, 0.8) & (0.8, 1.2) & (1.2, 2.4) & $>2.4$\\
%\hline
%               All           & 314 &  3879 &  9613 & 13071 & 12515 &  9071 &   625 \\
%             MS              & 188 &  3858 &  9444 & 12691 &  5357 &  1012 &    40 \\
%All ($N_{\rm visit}\geq 2$)     & 238 &  2564 &  5663 &  7088 &  6613 &  4220 &   172 \\
% MS ($N_{\rm visit}\geq 2$)   & 138 &  2551 &  5554 &  6866 &  2867 &   419 &     1 \\
%\hline\hline
%    \end{tabular}
%    \caption{The numbers of observed stars in different types. MS means the main sequence star. %\nvisitt{} stands for the number of stars which are visited more than twice and have at least a %\snrb{$\geq 10$} spectrum in each visit.}
%    \label{tab:nstar_type}
%\end{table}

\begin{table}
    \centering
    \begin{tabular}{cccccccc}
    \hline\hline
    TYPE  & O     & B           & A          & F         & G          & K          & M\\
    \bprp & $<-0.3$ & (-0.3, 0.1) &(0.1, 0.5)  &(0.5, 0.8) & (0.8, 1.2) & (1.2, 2.4) & $>2.4$\\
\hline
               All          & 314 &  3880 &  9610 & 13068 & 12511 &  9071 &   624 \\
             MS             & 188 &  3859 &  9441 & 12688 &  5354 &  1011 &    40 \\
All ($N_{\rm visit}\geq 2$)     & 245 &  2725 &  6203 &  7927 &  7280 &  4752 &   193 \\
 MS ($N_{\rm visit}\geq 2$)   & 143 &  2711 &  6091 &  7680 &  3083 &   462 &     2\\
\hline\hline
    \end{tabular}
    \caption{The numbers of observed stars in different types. MS means the main sequence star. \nvisitt{} stands for the number of stars which are visited more than twice and have at least a \snrb{$\geq 10$} spectrum in each visit. There are 49,078 sources that have parallax and \bprp{} of {\it Gaia}.}
    \label{tab:nstar_type}
\end{table}

\begin{table}
    \centering
    \begin{tabular}{cccccccc}
    \hline\hline
    TYPE  & O     & B           & A          & F         & G          & K          & M\\
    \bprp{} (mag) & $<-0.48$ & [-0.48, -0.03) &[-0.03, 0.33)  &[0.33, 0.71) & [0.71, 1.07) & [1.07, 2.28) & $\geq 2.28$\\
\hline
               All          & 174 &  1523 &  7794 & 13218 & 11641 & 13984 &   746 \\
             MS             & 53 &  1511 &  7716 & 12944 &  8883 &  1547 &    64 \\
All ($N_{\rm visit}\geq 2$)     & 140 &  1177 &  5377 &  8740 &  7329 &  8010 &   262 \\
 MS ($N_{\rm visit}\geq 2$)   & 41 &  1171 &  5314 &  8565 &  5562 &   843 &     8 \\
\hline\hline
    \end{tabular}
    \caption{The numbers of observed stars in different types. MS means the main sequence star. \nvisitt{} stands for the number of stars which are visited more than twice and have at least a \snrb{$\geq 10$} spectrum in each visit. There are 49,078 sources that have parallaxes and \bprp{} of {\it Gaia} EDR3.}
    \label{tab:nstar_type}
\end{table}

\subsection{Visit and repeat rates}
\begin{figure}
    \centering
    \subfloat{\includegraphics[width=0.46\textwidth]{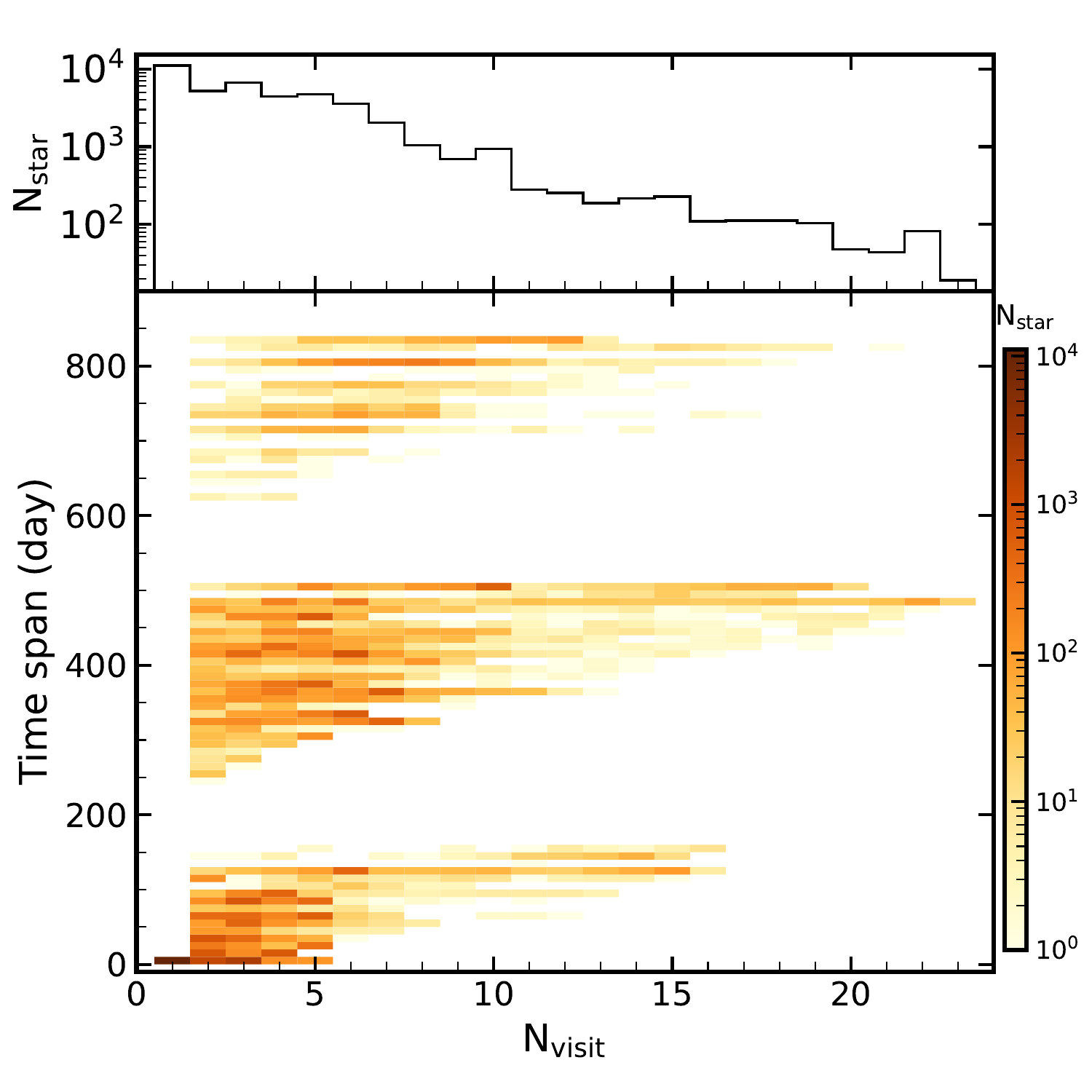}}(a)
    \subfloat{\includegraphics[width=0.46\textwidth]{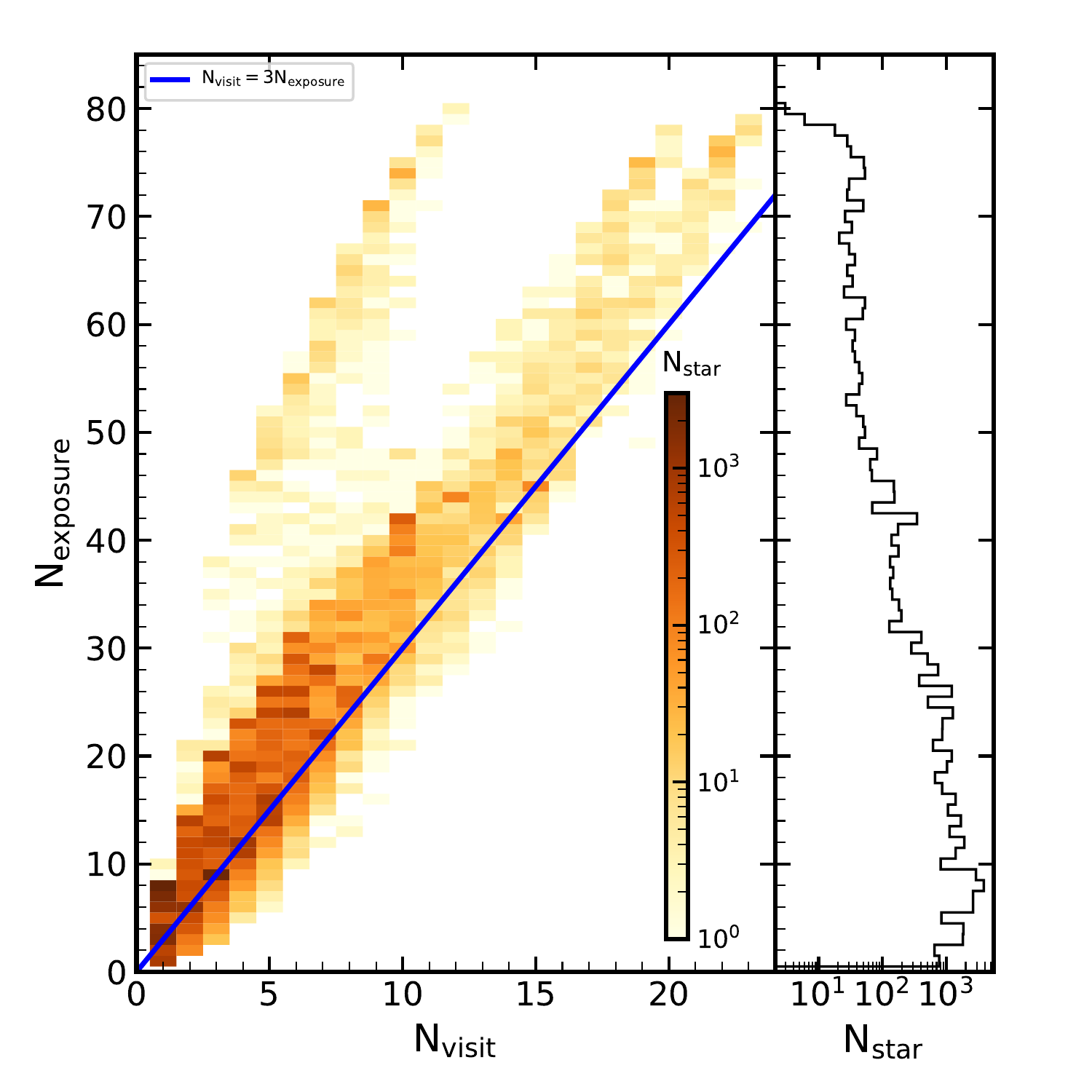}}(b) 
    %\makebox[\textwidth][c]{\includegraphics[width=0.7\textwidth]{figures/observe_visit.pdf}}
    \caption{Panel (a): the distribution of the number of visits and the time span of the stars observed in DR8 of MRS-B field, and there is at least a spectrum with \snrb{$\geq 10$} in each visit. Panel (a): the distribution of the numbers of visits and exposures, and the spectrum of each exposure has \snrb{$\geq 10$}. } %http://dr6.lamost.org/v2/doc/lr-data-production-description
    \label{fig:visit}
\end{figure}

We prefer to counts the stars that have at least a \snrb{$\geq 10$} spectrum in each visit, which would obtain reliable RVs. There are totally 42,193 sources with at least a \snrb{$\geq 10$} spectrum, and 31,073 sources are visited more than twice.
From Figure~\ref{fig:visit}, we can see that most of the stars were visited 2-5 times, and 2,735 sources have been visited more than 10 times. It hard to obtain good SNRs for all of the stars in each visit. For example, there are 1,713 sources in the TD114941N345554B01 plate that have been visited 15 times, while only 287 ($~17$\%) sources are visited 15 times with \snrb{$\geq 10$}. The stars that are observed with time span of $>700$ days came from the test plates of LAMOST-MRS pilot program such as HIP28117, which has the same center coordinates as TD055633N285632B01. Though we set three exposures per visit in the observation, a large fraction (28,589) of stars were observed with more exposures (see the left panel of Figure~\ref{fig:visit}) which profits from observations of the other sub-surveys. It would be useful to, but is not restricted to, study the binaries with periods less than one day.

%; 620 sources are visited $\geq14$ times owing to the high Galactic latitude plates (TD103949N392416B01 and TD114941N345554B01), see Table~\ref{tab:planid}

%%It should be noted that it is impossible to achieve good SNRs for all stars
%\begin{figure}
%    \centering
%    \makebox[\textwidth][c]{\includegraphics[width=0.7\textwidth]{figures/observe_visit.pdf}}
%    \caption{The distribution of the number of visits and the time span of the stars observed %in DR8 of MRS-B field. The stars that have at least a \snrb{$\geq 10$} spectrum in each %visit are counted.} %http://dr6.lamost.org/v2/doc/lr-data-production-description
%    \label{fig:visit}
%\end{figure}

\begin{figure}
    \centering
    \includegraphics[width=0.7\textwidth]{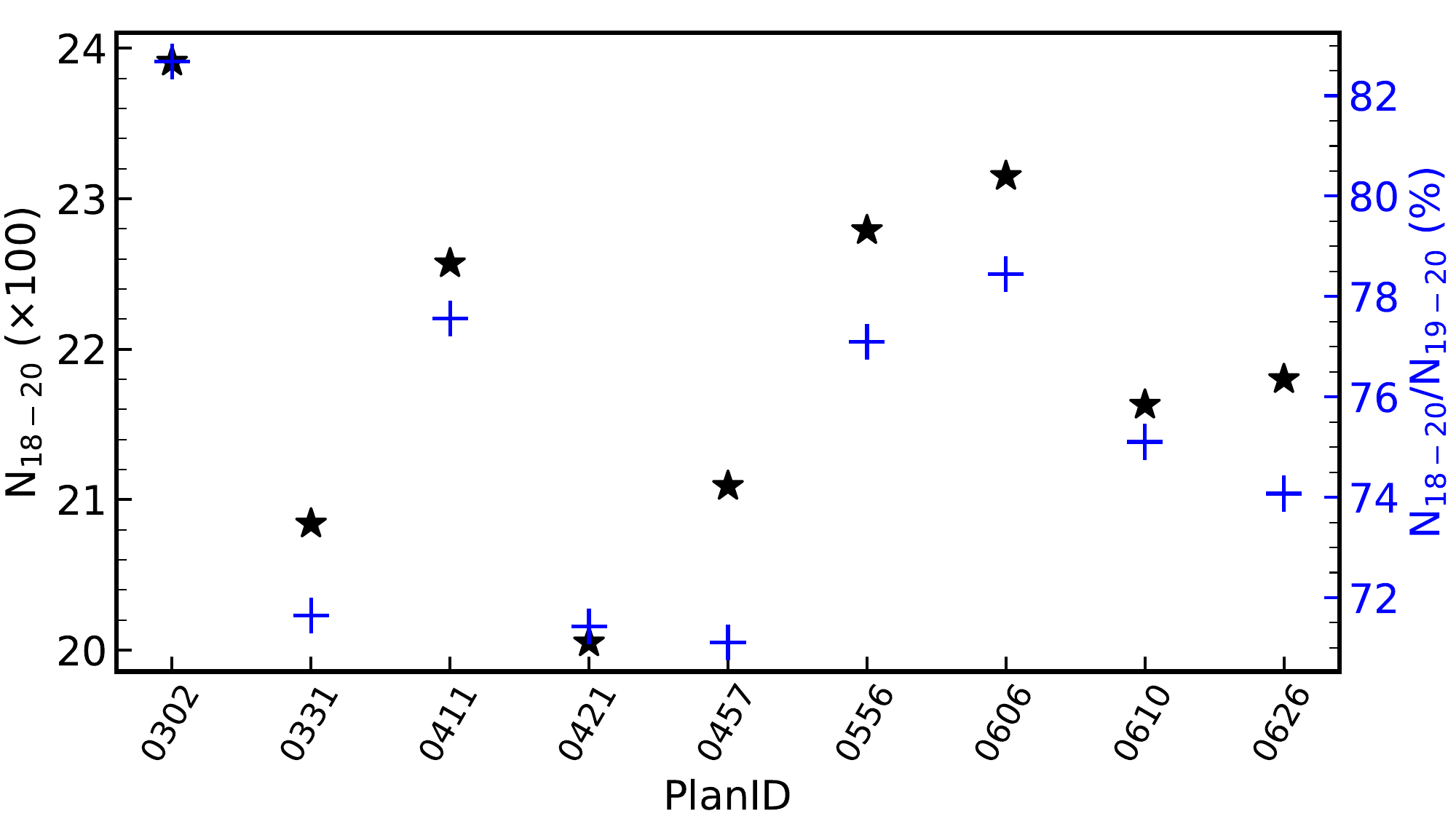}
    \caption{The repeat rate of plates observed from October 2018 to June 2020. $N_{\rm 18-19}$ and $N_{\rm 19-20}$ denote the number of star observed from October 2018 to July 2019 and October 2019 to June 2020, respectively. The detail number can be found in Table~\ref{tab:planid}.} %http://dr6.lamost.org/v2/doc/lr-data-production-description
    \label{fig:repeat}
\end{figure}

After the first year of observation of LAMOST-MRS, we keep 12 plates to observe and modified the observation priorities of their input sources. The repeat rate can be estimated by using the night plates observed in the second year. From Figure~\ref{fig:repeat}, we can see that more than 70\% of the sources are left for all plates. We lost 5,149 stars observed more than twice without considering the plate of TD043724N254338, which would be observed in the MRS-S field and has a similar priority to MRS-B field (the second and following years). The plate HIP28117 was visited five times from 2017-10-29 to 2017-11-08, and 1,747 sources were observed of which 516 stars are still observed in the TD055633N285632B01.

There are 3,4107 sources have been observed more than twice with \snrb{$>5$} or \snrr{$>5$} in MRS-B. We subtract the abandoned sources (5149+1231) which have been observed more than twice, and the remaining 27,727 sources will be observed in the next few years. There are still have 3 plates that are not observed in the second year (see, Table~\ref{tab:planid}). Two of them were only visited one time in the first year, so that 5,000 sources would be observed in the following years. We would obtain over 30,000 sources that have the potential to be visited more than 10 times. We can get more visits as considering the common sources that might be observed in the other sub-survey fields of LAMOST-MRS.

\subsection{Radial velocity}
\begin{figure}
    \centering
    \makebox[0.5\textwidth][c]{\includegraphics[width=0.7\textwidth]{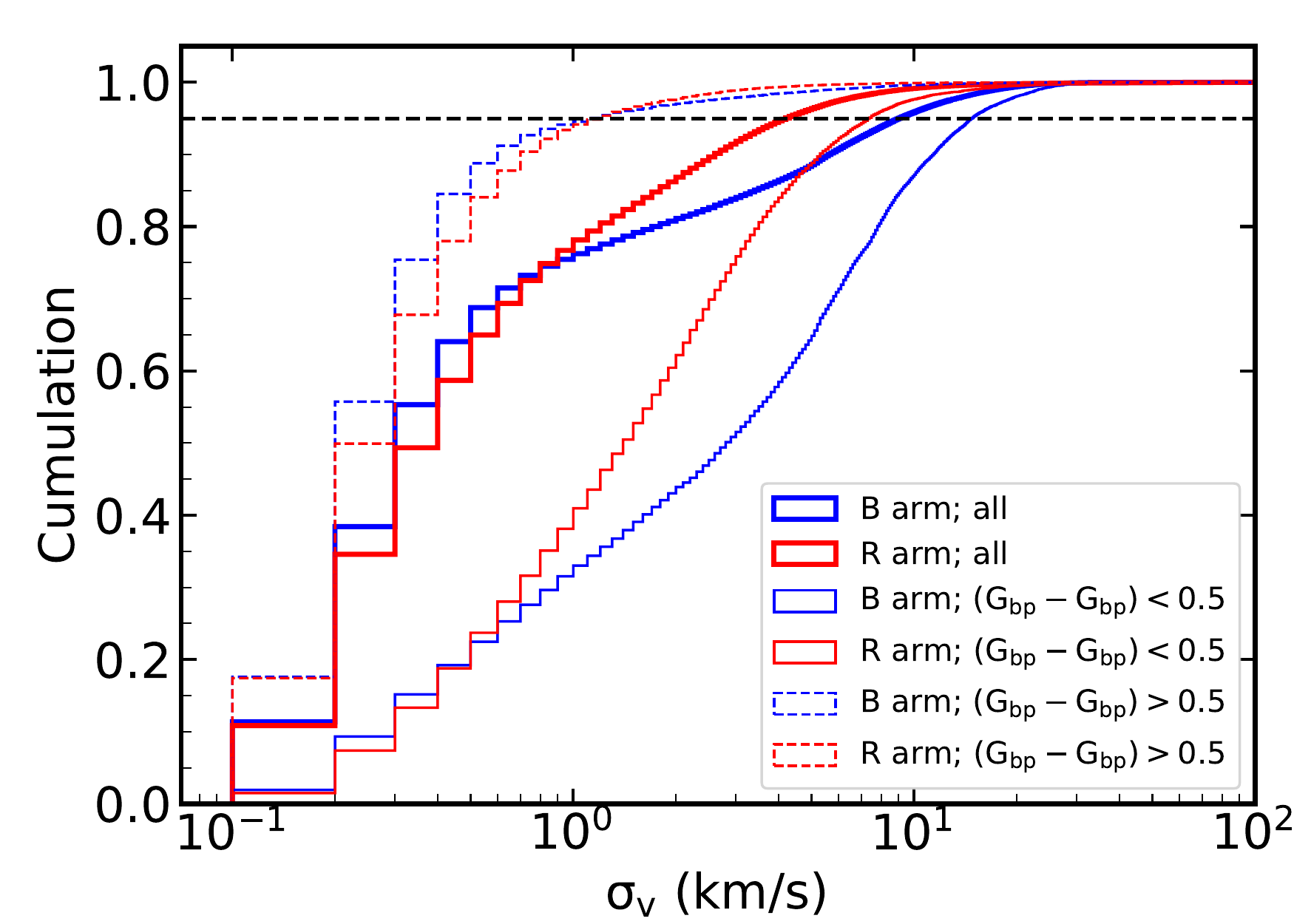}}
    \caption{The cumulative density function of RV uncertainties of the spectra with $SNR_{\rm B} \geq 10$. Blue and red colors stand for RV uncertainties measured from blue and red arm spectra, respectively. \bprp{} is corrected by using \textbf{dustmaps}. The black dashed line stands for 95\%.} %http://dr6.lamost.org/v2/doc/lr-data-production-description
    \label{fig:rverr}
\end{figure}

We measure the self-consistent stellar RV\cite{Zhangbo2021ApJS}, which reduces temporal variation of the RV zero-points (between exposures) by combining LAMOST-MRS with {\it Gaia} EDR3\cite{Gaiaedr32021A&A}. We find that $>95\%$ spectra with \snrb{$\geq 10$} have \sigmarv{$<10$} \kms in the MRS-B field. Comparing the cumulative density functions (CDF) of \sigmarvb{} and \sigmarvr{}, we find that CDF of \sigmarvb{} is steeper when \sigmarv{$<1$} \kms, as the thick lines of Figure~\ref{fig:rverr} shown, but CDF of \sigmarvr{} is steeper when \sigmarv{$>1$} \kms. Then, we divide the spectra into two sets by the corrected color \bprp{$=0.5$} of the stars. From the blue and red thin lines of Figure~\ref{fig:rverr}, we can see that CDF of \sigmarvr{} is steeper for early-type stars (\bprp{$<0.5$}), while the situation is inverse for the late-type stars (\bprp{$>0.5$}, see the dashed lines. It indicates that the RVs derived by the red arm spectra are preciser than the RVs of blue arm spectra for early-type stars due to that the absorption lines of early-type stars are weak and rare in the blue arm spectra. In addition, for the late-type stars, $>95\%$ spectra have \sigmarv{$<1$} \kms.

%We find that some of hot stars would be mistaken as poor metallicity stars. 
%We note that the LASP pipeline gives a relatively large scatter when it is applied to giant stars with a low surface gravity (Luo et al. 2015).

%\begin{figure}
%    \centering
%     \subfloat{\includegraphics[width=0.45\linewidth]{figures/cmd_feh.pdf}}
%     \hfill
%     \subfloat{\includegraphics[width=0.45\linewidth]{figures/kieldiagrm.pdf}}
%     %\bigskip
%     \subfloat{\includegraphics[width=0.45\linewidth]{figures/kieldiagrm.pdf}}
%     \hfill
%     \subfloat{\includegraphics[width=0.45\linewidth]{figures/kieldiagrm.pdf}}
%    \caption{The CMD and Kiel diagram colored by [Fe/H]. The stellar parameters are the median values from multiple measurements derived %     with the updated LASP pipeline.} %http://dr6.lamost.org/v2/doc/lr-data-production-description
%    \label{fig:cmd_kiel}
%\end{figure}

\section{Summary and Prospective}

After the first five years survey of LAMOST-LRS, LAMOST-MRS had been started regularly since 2018. We obtain about 20\% observation time to study the properties of close binary and exotic stars. We initially input 25 plates in the first year, and 16 plates were observed and 34 visits were obtained. It is difficult to observe the all plates with $>10$ visits in the next four years. So we dropped out 13 of them and added two new plates located at the high Galactic latitude. Checking the second year observation, we get 61 visits, of which 29 visits are from the two high Galactic latitude plates. We would achieve our goals to obtain about 30,000 sources with $N_{\rm vist} \geq 2$ by estimating the number of the visits during the 2-year observation. We might obtain more visits due to the common sources observed by the other sub-surveys of LAMOST-MRS.

%However, the number of stars that have \snrb{$\geq 10$} in each visit is  ratio only $\sim 17$. We might 

We place our plates a little away from the Galactic disk plane in order to pick out the exotic objects composited of WD+MS stars and reduce their influence on the statistic property of binary by using the UV catalog of GLAEX. We gave some identified exotic objects with the highest observation priorities, and DR8 include 347 the identified exotic objects in the MRS-B field. These objects could be well studied with LAMOST-MRS spectra. For example, we could study the variation of ${\rm H\alpha}$ emission line of ${\rm EB_{Xray}}$ and WR, or obtain orbital parameters of WD+FKG binaries. In addition, a few exotic objects ($\sim 1$\%) will not influence the process of analyzing the statistical properties of binary.

We also set high observation priorities for the low metallicity candidates. We found that there are about 200 sources with ${\rm [Fe/H]}<-1$ dex after checking DR8. It is possible to analyze the binary fraction of these metal-poor stars. For the source with $-0.6<{\rm [Fe/H]}<0.4$, we might be able to study the correlation between binary orbital parameters and metallicity abundance by using the all 5-year data, since there are $>1000$ sources in each of 0.2 dex bins. 

There are 53,360 sources in the MRS-B filed, of which 28,828 and 3,375 sources are visited $\geq 2$ and $\geq 10$ times, which have at least one \snrb{$\geq 10$} spectrum in each visit. We gave the brighter sources with higher observation priorities so that most of the observation stars belong to F-type, and majority of G- and K-type stars are located in the region of RC star. Each number of B-, A-, F-, and G-type dwarfs are more than 1500, these will give us an opportunity to obtain precise values of close binary fraction for the different type stars. By the way, the RV precision of early- and late-type stars are better than 1 and 10 \kms, respectively. Therefore, it becomes possible to study the binary statistic properties in smaller stellar mass bins and obtain more reliable relations between the orbital parameters. 

%%% RV and vists

In addition, we are also trying to use a few visits data of LAMOST-MRS to find exotic stars and study binarity. By using LAMOST-MRS DR7, we identified 1,162 Be stars\cite{Wangluqian2022ApJs} and a quadruple (2+2) system which is the most massive eclipse quadruple system ever found (in preparation, Li et al.). We found that the $f_{\rm b}$ vary from about 68\% to 44\% with stellar spectral type vary from B- to A-type\cite{Guoyanjun2022RAA}. The power law index of mass ratio distribution ($\gamma$) changes from $-0.42\pm0.27$ to $2.12\pm0.19$ when spectral type varies from A- to G-type by analyzing the SB2 stars~\cite{Lijiangdan2022ApJ}. We will resolve the orbital periods of the stars with the full 5-year observation and release them.

All the datasets presented in this paper, including the two catalogs and descriptions, are compiled in the Supplementary Materials.

\textbf{Acknowledgements:}
The authors thank Jian-Ning Fu, Xiao-Jie Xu, Yin-Bi Li, Zhong-Rui Bai and all of members of LAMOST-MRS Working Group for valuable discussions. 
This work is supported by the Natural Science Foundation of China (Nos. 12073070, 12173081, 12090043, 11873016, 12173013),
Yunnan Fundamental Research Project (No. 202101AV070001), the Science Research Grants from The China Manned Space Project (NOs. CMS-CSST-2021-A08, CMS-CSST-2021-A10, CMS-CSST-2021-B05)  and CAS `Light of West China' Program.

Guoshoujing Telescope (the Large Sky Area Multi-Object Fiber Spectroscopic Telescope LAMOST) is a National Major Scientific Project built by the Chinese Academy of Sciences. Funding for the project has been provided by the National Development and Reform Commission. LAMOST is operated and managed by the National Astronomical Observatories, Chinese Academy of Sciences.

This work has made use of data from the European Space Agency (ESA) mission \href{https://www.cosmos.esa.int/gaia}{\it Gaia}, processed by the {\it Gaia} Data Processing and Analysis Consortium (\href{https://www.cosmos.esa.int/web/gaia/dpac/consortium}{DPAC}). Funding for the DPAC has been provided by national institutions, in particular the institutions participating in the {\it Gaia} Multilateral Agreement.

The LAMOST FELLOWSHIP is supported by Special Funding for Advanced Users, budgeted and administrated by Center for Astronomical Mega-Science, Chinese Academy of Sciences (CAMS). This work is supported by Cultivation Project for LAMOST Scientific Payoff and Research Achievement of CAMS-CAS and Special research assistant program of Chinese Academy of Sciences.

{\it Software}: python, laspec\cite{Zhangbo2020ApJS}, astropy\cite{Astropy2018AJ}, \href{http://www.star.bris.ac.uk/~mbt/topcat/}{TOPCAT}\cite{Topcat2005ASPC}.

\end{CJK*}  %% end the Chinese environment

\bibliography{lamostbinary.bib}
\bibliographystyle{cpb}

\end{document}